\def\sphinxdocclass{article}
\documentclass[A4paper,12pt,english]{sphinxhowto}

\usepackage{fixltx2e} 
\usepackage{cmap} 
\usepackage{ifthen}                                                                                                                           
\usepackage[T1]{fontenc}                                                                                                                      
\usepackage[utf8]{inputenc}                                                                                                                   
\usepackage[pdftex]{graphicx}     
\usepackage{color}             
                                                                                                                                                                                                                                                             
\usepackage{mathptmx} 
\usepackage[scaled=.90]{helvet}                                                                                                               
\usepackage{courier}                                                                                                                          

\usepackage{babel}
\usepackage{amsfonts}
\usepackage{amssymb,float}
\usepackage{xspace}
\usepackage{placeins}
\usepackage{rotating}
\usepackage{dcolumn}
\usepackage{bm}
\usepackage{cite}
\usepackage{amsmath}
\usepackage{indentfirst} 
\usepackage{tikz}
\usepackage{longtable,multirow}
\usetikzlibrary{shapes,arrows}
\usepackage{sphinx}

\usepackage{times}
\usepackage[Bjarne]{fncychap}
\usepackage{eqparbox}

\usepackage{blindtext}
\usepackage{enumitem}

\DeclareUnicodeCharacter{00A0}{\nobreakspace}

\def\eg{{e.g.}}
\def\ie{{i.e.}}

\def\code#1{{\tt{#1}}}

\def\lsim{\mathrel{\raise.3ex\hbox{$<$\kern-.75em\lower1ex\hbox{$\sim$}}}}
\def\gsim{\mathrel{\raise.3ex\hbox{$>$\kern-.75em\lower1ex\hbox{$\sim$}}}}
\def\ifmath#1{\relax\ifmmode #1\else $#1$\fi}

\def\smodels{{SModelS\,v1.1}}
\def\smodelsnn{{SModelS}}

\newcommand{\amc}{{\sc MadGraph5\textunderscore}a{\sc MC@NLO}}

\newcommand{\pythiaE}{{\sc Pythia\,8}\xspace}
\newcommand{\pythiaET}{{\sc Pythia\,8.2}\xspace}
\newcommand{\pythiaS}{{\sc Pythia\,6}\xspace}
\newcommand{\pythiaSF}{{\sc Pythia\,6.4}\xspace}
\newcommand{\pythiaSFTS}{{\sc Pythia\,6.4.27}\xspace}
\newcommand{\pythia}{{\sc Pythia}\xspace}
\newcommand{\nllfast}{{\sc NLLfast}\xspace}

\newcommand{\delphes}{{\sc Delphes\,3}}
\newcommand{\MA}{{\sc MadAnalysis\,5}}
\newcommand{\CM}{{\sc CheckMATE}}

\makeatletter
\def\PYG@reset{\let\PYG@it=\relax \let\PYG@bf=\relax%
    \let\PYG@ul=\relax \let\PYG@tc=\relax%
    \let\PYG@bc=\relax \let\PYG@ff=\relax}
\def\PYG@tok#1{\csname PYG@tok@#1\endcsname}
\def\PYG@toks#1+{\ifx\relax#1\empty\else%
    \PYG@tok{#1}\expandafter\PYG@toks\fi}
\def\PYG@do#1{\PYG@bc{\PYG@tc{\PYG@ul{%
    \PYG@it{\PYG@bf{\PYG@ff{#1}}}}}}}
\def\PYG#1#2{\PYG@reset\PYG@toks#1+\relax+\PYG@do{#2}}

\def\PYGZus{\char`\_}

\def\PYGZsh{\char`\#}

\def\PYGZhy{\char`\-}
\def\PYGZsq{\char`\'}

\makeatother

\renewcommand\PYGZsq{\textquotesingle}

\begin{document}

\thispagestyle{empty}

\begin{center}


\includegraphics[width=8cm]{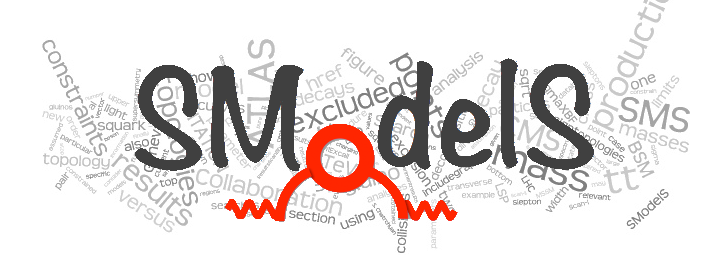}

\vspace*{4mm}

{\LARGE\bf SModelS v1.1 user manual} \\[2mm]
{\bf --- improving simplified model constraints with efficiency maps ---}



\vspace*{8mm}

\renewcommand{\thefootnote}{\fnsymbol{footnote}}

{\normalsize 
Federico~Ambrogi,$^{1}$
Sabine~Kraml,$^{2}$  
Suchita~Kulkarni,$^{1}$  
Ursula~Laa,$^{2,3}$  
Andre~Lessa,$^{4}$  
Veronika~Magerl,$^{5}$  
Jory~Sonneveld,$^{6}$ 
Michael~Traub,$^{\,1}$\footnote{Present address: Kirchstrasse 2a, 79312 Emmendingen, Germany}  
Wolfgang~Waltenberger$^{\,1}$  
} 

\renewcommand{\thefootnote}{\arabic{footnote}}

\vspace*{4mm} 

{\small \it 
$^1\,$Institut f\"ur Hochenergiephysik,  \"Osterreichische Akademie der Wissenschaften,\\ Nikolsdorfer Gasse 18, 1050 Wien, Austria\\[1mm]
$^2\,$Laboratoire de Physique Subatomique et de Cosmologie, 
Universit\'e Grenoble-Alpes, CNRS/IN2P3,\\ 53 Avenue des Martyrs, F-38026 Grenoble, France\\[1mm]
$^3\,$ LAPTh, Universit\'e Savoie Mont Blanc, CNRS, B.P.110 Annecy-le-Vieux,\\ F-74941 Annecy Cedex, France\\[1mm]
$^4\,$Centro de Ci\^encias Naturais e Humanas, Universidade Federal do ABC,\\ Santo Andr\'e, 09210-580 SP, Brazil\\[1mm]
$^5\,$Fakult\"at f\"ur Mathematik und Physik, Albert-Ludwigs-Universit\"at, 79104 Freiburg, Germany\\[1mm]
$^6\,$Institut f\"ur Experimentalphysik, Universit\"at Hamburg, 22761 Hamburg, Germany
}

\vspace*{4mm}

Email: smodels-users@lists.oeaw.ac.at

\vspace*{4mm}

\begin{abstract}
SModelS is an automatised tool for the interpretation of simplified model results from the LHC. 
It allows to decompose models of new physics obeying a $\mathbb{Z}_2$
symmetry into simplified model components, and 
to compare these against 
a large database of experimental results. 
The first release of SModelS, v1.0, 
used only cross section upper limit maps provided by the experimental collaborations.
In this new release, v1.1, we extend the functionality of SModelS to efficiency maps.  
This increases the constraining power of the software, 
as efficiency maps allow to combine 
contributions to the same signal region from different simplified models. 
Other new features of version~1.1 include likelihood and $\chi^2$ calculations, extended information on the topology coverage, 
an extended database of experimental results as well as major speed upgrades for both the code and the database. 
We describe in detail the concepts and procedures used in \smodels, 
explaining 
in particular 
how upper limits and efficiency map results are dealt with in parallel.
Detailed instructions for code usage are also provided.
\end{abstract}

\end{center}

\setcounter{page}{0}\clearpage 

\tableofcontents

\setcounter{page}{1}
\pagestyle{plain}

\clearpage
\section{Introduction}\label{sec:intro}

The ATLAS and CMS experiments at the Large Hadron Collider (LHC) are searching for new physics beyond the Standard Model (BSM)
in many different channels.
An important class of these searches considers final states with large missing transverse energy (MET), 
targeting supersymmetry (SUSY) with R-parity conservation or other models with a new conserved parity. 
In order to design optimal analyses that look for specific final states and to communicate the results,  
the concept of simplified models~\cite{hep-ph/0703088, 0810.3921, 1105.2838, Okawa:2011xg, 1301.2175} 
has been widely adopted by the experimental collaborations. 

While limits in terms of simplified model spectra (SMS) are a convenient way to
illustrate and compare the reach of specific analyses, understanding how they
constrain a realistic model with a multitude of relevant production channels and
decay modes quickly becomes a non-trivial task.
To tackle this issue we have created
\smodelsnn~\cite{Kraml:2013mwa,Kraml:2014sna},
an automatised tool for interpreting simplified model results from the LHC.
The principle of \smodelsnn\ is to decompose BSM collider signatures featuring a
$\mathbb{Z}_2$ symmetry into simplified model topologies, using a generic
procedure where each SMS is defined by the vertex structure and the SM final
state particles; BSM particles are described only by their masses, production
cross sections and branching ratios.

The SModelS decomposition corresponds to an approximate mapping of the
original model signatures to a coherent sum of simplified model topologies. 
The underlying assumption~\cite{Kraml:2013mwa} is that differences in the event kinematics 
(\eg\ from different production mechanisms or from the spin of the BSM particle)
do not significantly affect the signal selection efficiencies. 
The tool can then be used for any BSM model with a
$\mathbb{Z}_2$-like symmetry as long as all heavier R-odd particles (cascade-)decay
promptly to the lightest one, which should be electrically and colour neutral.%
\footnote{Charged tracks may also be treated in an SMS context~\cite{Heisig:2015yla}
and will be available in future versions of \smodelsnn.} 
Note that due to the $\mathbb{Z}_2$ symmetry only pair production
or associated production of two BSM particles is considered, and MET is always
implied in the final state description.
Regarding experimental results, the tool contains a large database of SMS results from ATLAS and CMS SUSY searches.

Since the publication of SModelS\,v1.0~\cite{Kraml:2014sna} in December
2014, the code base has undergone significant structural changes. Version~1.1
comes with many new features, the most important one being the {\it extension to efficiency maps}. 

\noindent
The advantage of efficiency maps (EM) over the previously (and still) used cross
section upper limit (UL) maps is that they allow to combine contributions from
different SMS topologies to the same signal region (SR) of a given experimental
analysis. This significantly enhances the coverage and constraining power of the
SMS results.
Further novelties of this release include:

\begin{description} 
  \item [\quad --] {\it a new and more flexible database format;}
  \item [\quad --] {\it extended information on the topology coverage;}
  \item [\quad --] {\it inclusion of likelihood and \(\chi^2\) calculation for {EM-type
  results};}
  \item [\quad --] {\it inclusion of a database browser tool for easy access to the
  information stored in the database;}
  \item [\quad --] {\it performance improvement for the decomposition of the input model;}
  \item [\quad --] {\it inclusion of new simplified model results to the database,
  including a few 13 TeV results;}
  \item [\quad --] {\it overall significant speedups of the code and the database.}
\end{description}

The purpose of this paper is to present these new features, 
describing in detail the concepts and procedures used in the code and the database. 
Particular attention is given to how upper limits and efficiency maps are
dealt with simultaneously.

The (re-)interpretation of LHC searches, which is the main goal of \smodelsnn,
has recently become a very active field.
Several other public tools have been developed by different groups. 
In particular, FastLim~\cite{Papucci:2014rja} and XQCAT~\cite{Barducci:2014gna}
employ pre-computed efficiency maps to constrain, respectively, the minimal
supersymmetric standard model (MSSM) and extra heavy quark scenarios.
SUSY-AI~\cite{Caron:2016hib} is a code which has been trained with machine
learning techniques to reproduce the ATLAS exclusion for the phenomenological
MSSM.
Finally, \CM~\cite{Drees:2013wra,Dercks:2016npn}, \MA~\cite{Conte:2014zja,Dumont:2014tja}, Rivet ($\ge2.5$) \cite{Buckley:2010ar} 
and GAMBIT's ColliderBit\cite{Athron:2017ard,Balazs:2017moi} 
allow for more general recasting of ATLAS and CMS searches based on Monte Carlo event simulation.  
Last but not least, Contur~\cite{Butterworth:2016sqg} aims at constraining BSM
scenarios from differential Standard Model measurements based on the
Rivet~\cite{Buckley:2010ar} toolkit. 

The rest of the paper is organised as follows.
We start in Section~\ref{SModelSDefs:basic-concepts-and-definitions} by explaining the basic concepts of \smodelsnn, 
including the structure of UL-type and EM-type results and of the database of experimental results. 
In Section~\ref{Structure:smodels-structure}, we then discuss in detail the decomposition procedure, the computation of the theory predictions, the comparison of theory predictions against the experimental results, and how missing topologies are identified. 
How to run \smodelsnn\ is shown in Section~\ref{RunningSModelS:running-smodels}. 
Section~\ref{Conclusions} contains conclusions and an outlook.

Installation instructions are provided in Appendix~\ref{Installation}.
Appendix~\ref{Tools:smodels-tools} describes some useful SModelS-internal tools, 
while Appendix~\ref{EMcreation} provides information about 
`home-grown' efficiency maps included in the database.
A complete, browsable manual and an extensive code documentation are provided in
html format in the `docs' folder of the distribution; they are also available
online at \cite{smodels:wiki}.

In case of problems using \smodelsnn, we kindly ask the user to contact the
authors via \code{smodels-users@lists.oeaw.ac.at}.\\

\vspace*{6mm}

\hrule
\vspace*{1mm}
\noindent {\bf Note:}

When using \smodels\ for scientific work (or any other purposes), please cite this paper 
as well as the original SModelS publication~\cite{Kraml:2013mwa}.
In scientific publications, please also cite
the programs \smodels\ makes use of: \pythiaSF~\cite{Sjostrand:2006za} and/or
\pythiaET~\cite{Sjostrand:2006za,Sjostrand:2014zea}, \nllfast
\cite{Beenakker:1996ch,Beenakker:1997ut,Kulesza:2008jb,Kulesza:2009kq,Beenakker:2009ha,Beenakker:2010nq,Beenakker:2011fu},
and PySLHA~\cite{Buckley:2013jua}.
For convenience, these citations are provided as a bibtex file in the
\smodelsnn\ distribution.

We also strongly recommend to cite all the experimental results used; a separate
bibtex file is provided to that effect in the \smodelsnn\ database folder. In case
you use the FastLim efficiency maps in \smodelsnn\ (see `adding FastLim data' in Appendix~\ref{Tools:smodels-tools}),
please also properly cite FastLim~\cite{Papucci:2014rja} and the relevant
experimental results.  
These references are also provided in a bibtex file with the distribution.

\vspace*{2mm}

\hrule

\section{Basic Concepts and Structure}
\label{SModelSDefs:basic-concepts-and-definitions}

A central task of \smodelsnn\ is the decomposition of a complex BSM model into
its set of simplified model
topologies. As mentioned in the introduction, this decomposition relies on 
several assumptions~\cite{Kraml:2013mwa}:
first, the production channel is not taken into account, and only on-shell
particles are considered in cascade decays.
Virtual particles are replaced by an effective vertex, where only the on-shell
decay products are specified.
Additionally, new states are described only by their masses, neglecting all other quantum numbers, 
thus 
disregarding 
potential influences of the 
spin and coupling structure on the selection efficiencies.
Finally it should be noted that the SMS approach is only valid within the narrow
width approximation.
For a safe application of \smodelsnn\ (in particular to non-MSSM or non-SUSY
scenarios), these assumptions should be understood and, if needed, verified.

A few studies on the validity of the SMS assumptions are available.
For example, the effects of alternative production channels in squark simplified
models were studied in~\cite{Edelhauser:2014ena}. The effect of a different spin
structure was studied for the case of the dijet+MET final state
in~\cite{Edelhauser:2015ksa}, for the dilepton+MET final state
in~\cite{Arina:2015uea} and for $t\bar{t}$+MET final states
in~\cite{Kraml:2016eti}. 
For all these cases it was found that the application of SMS limits is safe.
Generally, however, the validity of the SMS assumptions will depend on the
concrete model under consideration, as well as details of the experimental
search. In particular, an inclusive cut-and-count search might be less sensitive
to differences than a shape-based analysis. Examples for which the SMS 
assumptions do not hold include the mono-X analyses performed in the context of
dark matter searches. Dark matter simplified model results are therefore not
included in \smodels.

In this section we first explain the basic concepts and language used in \smodels\ 
to describe simplified model topologies. This is followed by detailed descriptions 
of UL-type and EM-type results and of the database of experimental results.

\subsection{Simplified Model Definitions}
\label{TheoryDefinitions:theorydefs}\label{TheoryDefinitions::doc}\label{TheoryDefinitions:theory-definitions}

\subparagraph{Elements}
\label{TheoryDefinitions:elements}\label{TheoryDefinitions:element} 

\begin{figure}[h!]
	\begin{center}
		\includegraphics[width=0.44\linewidth]{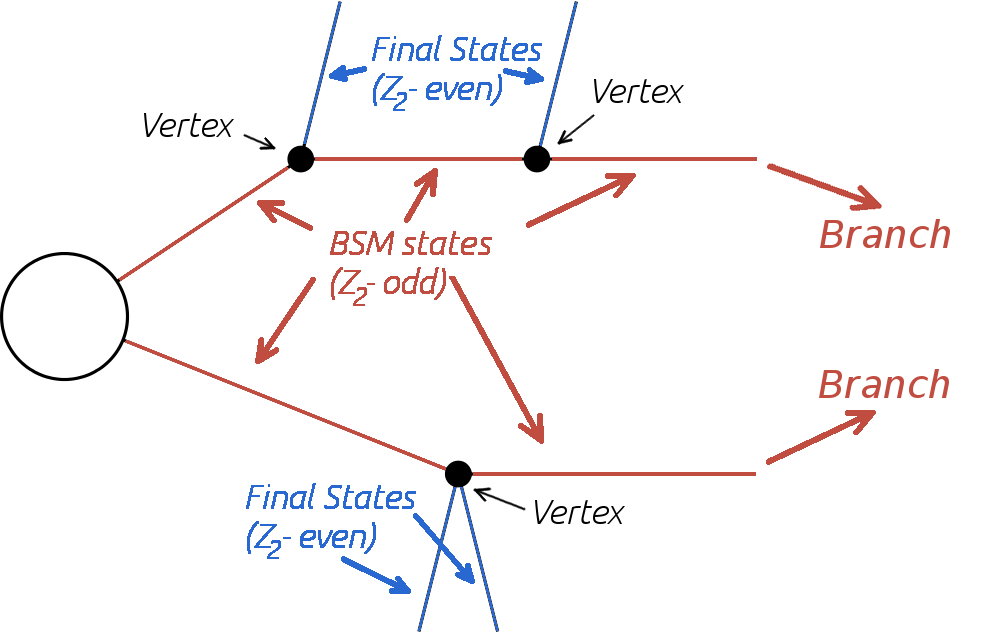}\qquad\includegraphics[width=0.34\linewidth]{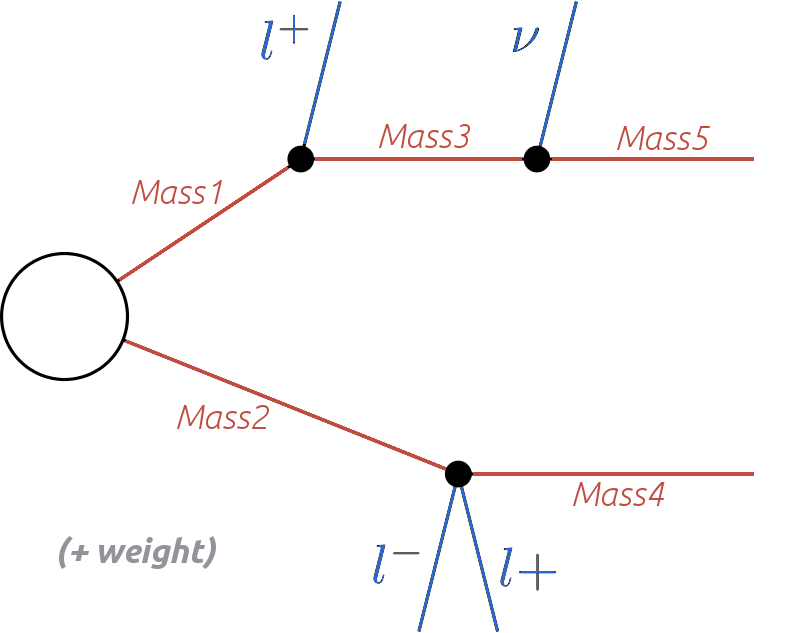}
	\end{center}
\caption{Representation of an element, including its overall properties (left)
and a concrete example with the information used in
\smodelsnn\ (right).}\label{TheoryDefinitions:elementscheme}
\end{figure}

A simplified model topology representing a specific cascade decay of a pair of
BSM states produced in the hard scattering is called an `element' in the SModelS
language. Elements contain the final states ($\mathbb{Z}_2$-even) particles
appearing in the cascade decay as well as the masses of the BSM
($\mathbb{Z}_2$-odd) states 
in the decay chain. 
An element may also hold information about its corresponding `weight'
(cross section times branching ratio times efficiency).\footnote{In order to
treat the UL and EM map results on the same footing, \smodels\ applies a trivial
binary efficiency to elements for UL-type results as will be explained later.} 
Figure~\ref{TheoryDefinitions:elementscheme} illustrates an element and its properties.

\smodelsnn\ works under the inherent assumption that, for collider purposes,
all the essential pro\-per\-ties of a BSM model can be encapsulated by its
elements. While some caveats apply (see above), 
such an assumption is extremely helpful to cast the theoretical predictions of a
specific BSM model in a model-independent framework, which can then be compared
against the corresponding experimental limits.
For instance, as shown in Figure~\ref{TheoryDefinitions:elementscheme}, only the
masses of the BSM states are used; other properties, such as their spins or
other quantum numbers are ignored (the PID's are, however, stored for book-keeping).

Below we describe in more detail the element properties and their implementation
in SModelS.

\subparagraph{Vertices}
\label{TheoryDefinitions:vertex}\label{TheoryDefinitions:vertices}

Each $\mathbb{Z}_2$-odd decay is represented by a vertex containing its final
states (one $\mathbb{Z}_2$-odd state and the $\mathbb{Z}_2$-even particles). 

\subparagraph{Final States ($\mathbb{Z}_2$-even)}
\label{TheoryDefinitions:final-states-z2-even}\label{TheoryDefinitions:final-states} 

Final states indicate all $\mathbb{Z}_2$-even states coming out of a vertex. In most cases,
these correspond to Standard Model particles (leptons, gauge bosons, Higgs,...).
Note that, if the input model contains BSM states which are $\mathbb{Z}_2$-even, 
such as additional Higgs bosons, these are also treated as final states. In
contrast, stable or long-lived $\mathbb{Z}_2$-odd particles which might appear
in the detector (either as MET or charged tracks) are \emph{not} classified as
final states.

\subparagraph{Intermediate States ($\mathbb{Z}_2$-odd)}
\label{TheoryDefinitions:odd-states}\label{TheoryDefinitions:intermediate-states-z2-odd}

The $\mathbb{Z}_2$-odd states are always assumed to consist of BSM particles
with $\mathbb{Z}_2$ conserving decays of the form: ($\mathbb{Z}_2$-odd state)
\(\rightarrow\)  ($\mathbb{Z}_2$-odd state) +
final states. The
only information kept from the intermediate states are their masses, see Figure~\ref{TheoryDefinitions:elementscheme} (right). 
If an intermediate state is stable and neutral, it is considered as a MET signal.


\subparagraph{Branches}
\label{TheoryDefinitions:branches}\label{TheoryDefinitions:branch}

\begin{figure}[h!]
	\begin{center}
	\includegraphics[width=0.350\linewidth]{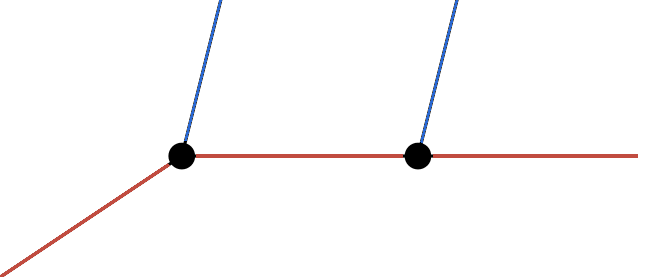} \qquad
	\includegraphics[width=0.350\linewidth]{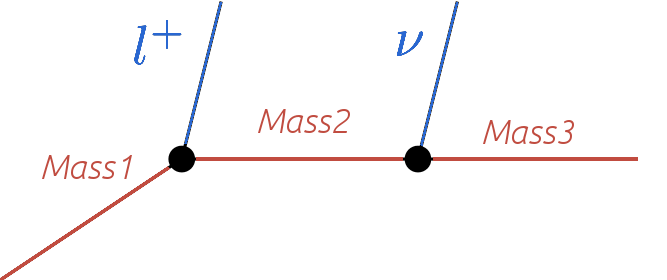}
	\end{center}
\caption{Representation of a
branch, including its overall structure
(left) and a concrete example with all the information it holds in the SModelS
scheme (right).}\label{TheoryDefinitions:branchTop}
\end{figure}

A branch is the basic substructure of an element. It represents the cascade decays of a single initial $\mathbb{Z}_2$-odd state. The diagram
in Figure~\ref{TheoryDefinitions:branchTop}  illustrates an example of a branch.

The structure of each branch is fully defined by its number of vertices and the
number of final states coming out of each vertex. Furthermore,  the branch also holds the
information about 
the final states originating from each vertex and the masses of the
intermediate states, as shown 
in Figure~\ref{TheoryDefinitions:branchTop} (right).

\subparagraph{Element Representation: Bracket Notation}
\label{TheoryDefinitions:element-representation-bracket-notation}\label{TheoryDefinitions:notation}

\begin{figure}[h!]
	\begin{center}
	\includegraphics[width=0.7\linewidth]{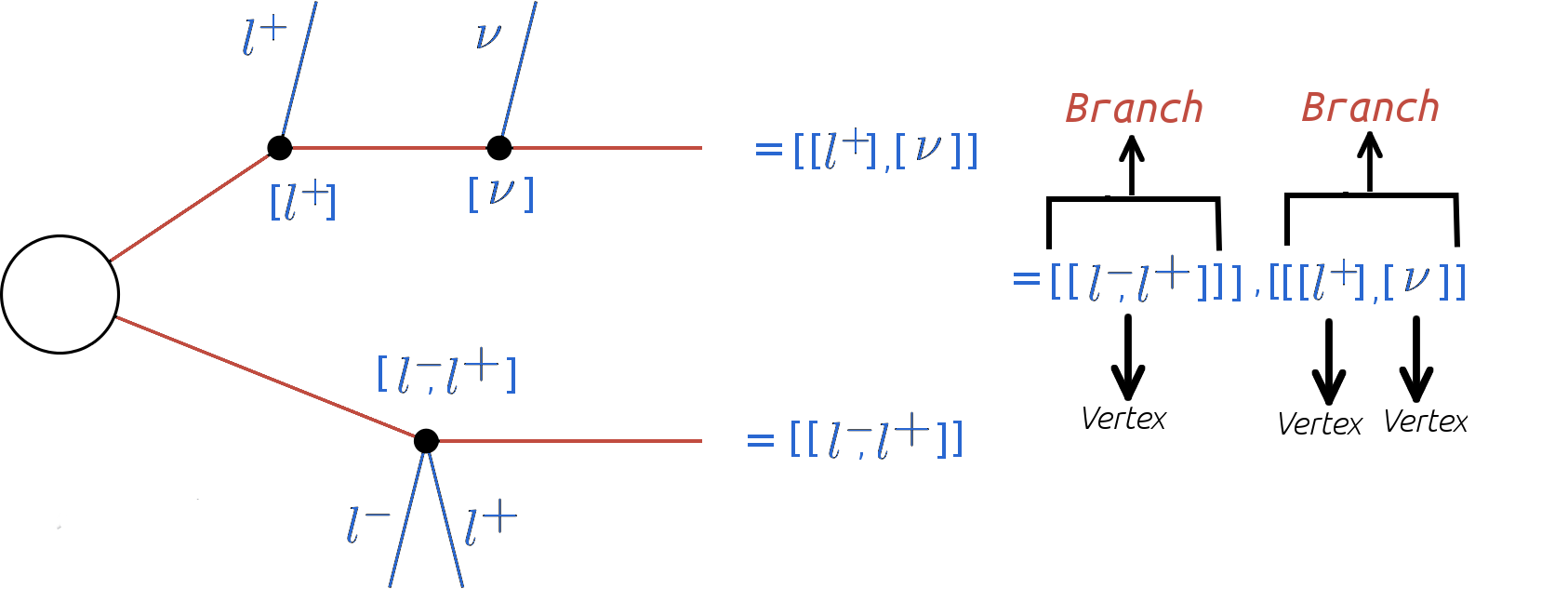}
	\end{center}
\caption{Conversion of an element to the bracket notation used in SModelS.}
\label{TheoryDefinitions:bracketnotation}
\end{figure}

The structure and final states of elements can conveniently be represented in
textual form using a notation of nested brackets.
Figure~\ref{TheoryDefinitions:bracketnotation} shows how to convert between the
graphical and bracket representations of an element.

\begin{figure}[t!]
	\begin{center}
	\includegraphics[width=0.75\linewidth]{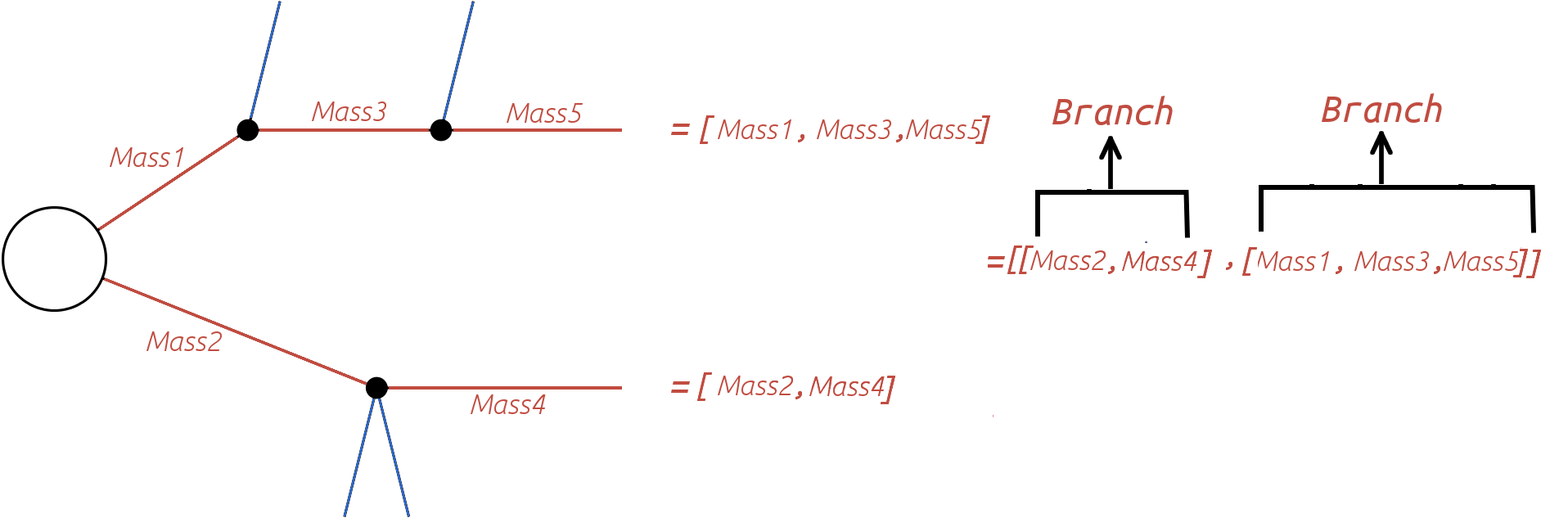}
	\end{center}
\caption{Example of an element `mass array', as used by SModelS.}
\label{TheoryDefinitions:massnotation}
\end{figure}

The brackets are ordered and nested in the following way. The outermost brackets
correspond to the branches of the
element. The branches are sorted
according to their size and each
branch contains an \emph{ordered} list of vertices.
Each vertex contains a list of the
final states originating from it. Schematically, for the example in
Figure~\ref{TheoryDefinitions:bracketnotation}, we
have:

\begin{Verbatim}[commandchars=\\\{\},frame=lines]
element = [branch1, branch2]
   branch1 = [vertex1]
      vertex1 = [l+,l\PYGZhy{}]
   branch2 = [vertex1,vertex2]
      vertex1 = [l+]
      vertex2 = [nu]
\end{Verbatim}

Using the above scheme it is possible to unambiguously describe each
element with a simple list of
nested brackets. However, in order to fully specify all the information relative
to a single element, we must also
include the list of
intermediate state
masses and the element weight. The
intermediate state
masses are represented by a mass array for each branch, as shown 
in Figure~\ref{TheoryDefinitions:massnotation}.

\subparagraph{Topologies}
\label{TheoryDefinitions:topologies}\label{TheoryDefinitions:topology}

\begin{figure}[t!]
	\begin{center}
	\includegraphics[width=0.33\linewidth]{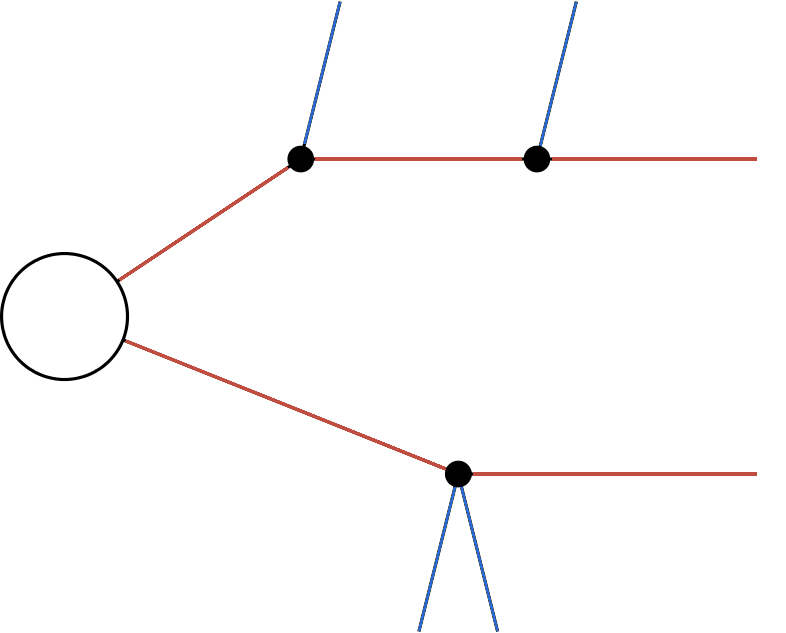}
	\end{center}
\caption{An example of a topology, which represents a class of elements.}
\label{TheoryDefinitions:globTop}
\end{figure}

It is often useful to classify
elements according to their
overall structure or topology. Each topology corresponds to an
\emph{undressed} element, removed
of its final states
and $\mathbb{Z}_2$-odd masses. Therefore the topology is fully determined
by its number of branches, number of vertices in each
branch and number of
final states coming
out of each vertex. An example of
a  topology is shown in Figure~\ref{TheoryDefinitions:globTop}.

Within SModelS, elements are grouped according to their topology. Hence 
topologies represent a list of elements sharing a common basic structure (same
number of branches, vertices and final states in each vertex).

\subsection{Database Definitions and Structure}
\label{DatabaseDefinitions:databasedefs}\label{DatabaseDefinitions::doc}\label{DatabaseDefinitions:database-definitions}

SModelS contains a large database of SMS results from ATLAS and CMS SUSY searches.\footnote{Extensions to non-MET searches are foreseen for future SModelS releases.} 
Starting with version~1.1, two types of experimental constraints are used: 
\begin{itemize}
	\item {\bf Upper Limit (UL)} constraints: constrains on \(\sigma \times BR\) of
	simplified models, usually provided by the experimental collaborations;  
	\item {\bf Efficiency Map (EM)} constraints: constrains the total signal
	(\(\sum \sigma \times BR \times \epsilon\)) in a specific signal
	region. Here \(\epsilon\) denotes the acceptance times efficiency.
	These are either directly provided by the experimental collaborations or computed by
	theory groups.  
\end{itemize}

Although the two types of constraints above are very distinct, both the folder
structure and the object structure of SModelS are sufficiently flexible to
simultaneously handle both UL-type and EM-type results. 
The database itself can be seen as a collection of experimental results,
each of which obey the following hierarchical structure:

\begin{figure}[h!t]\centering
	\includegraphics[width=0.98\linewidth]{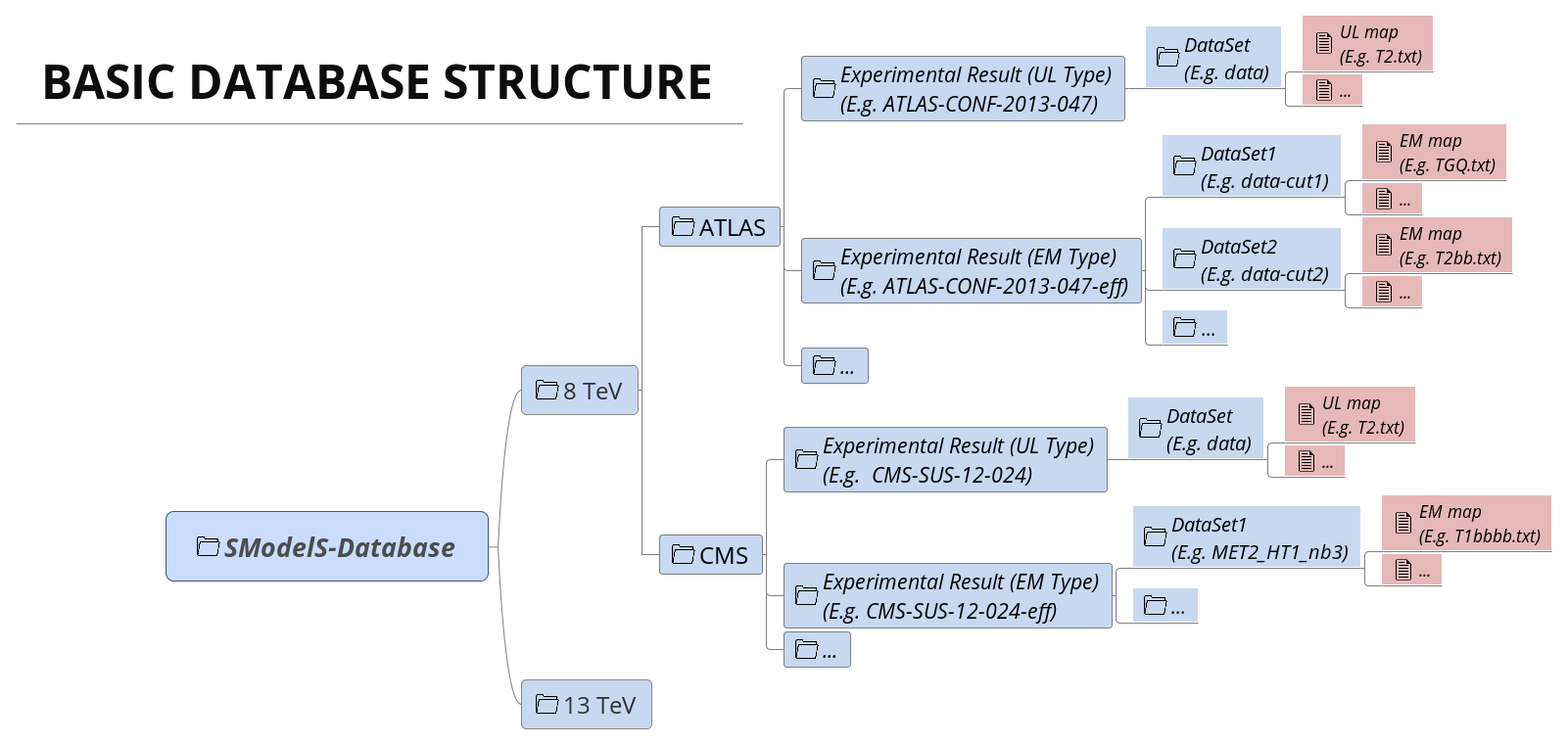}
\caption{Basic folder structure of the SModelS database.}
\label{DatabaseDefinitions:databasescheme}
\end{figure}

\begin{samepage}
	\begin{itemize}
	\item {\bf Experimental Result: } corresponds to an experimental publication (published paper or preliminary result like a conference note or analysis summary) 
	and contains a list of Data Sets as well as general
	information about the result (luminosity, publication reference, ...).
	
		\begin{itemize}
		\item {\bf Data Set: } 
		represents a distinct signal region for a given experimental result. 
		For EM-type results, there is one data set per signal region, each containing the relevant efficiency maps. 
		UL-type results are treated analogously, but, since they correspond to the best signal region or a combination of signal regions, they have just one data set containing the relevant upper limit map. 

			\begin{itemize}
			\item {\bf Upper Limit map:} contains the upper limit constraints for
			UL-type results. Each map
			refers to a single simplified model (or more precisely to a single
			element or sum of
			elements).

			\item {\bf Efficiency map:} contains the efficiencies for
			EM-type results. Each map
			refers to a single element or sum of elements.

			\end{itemize}

		\end{itemize}

	\end{itemize}
\end{samepage}

A schematic summary of the above structure is shown in Figure~\ref{DatabaseDefinitions:databasescheme}.
Below we describe in more detail the main concepts and building blocks of the
SModelS database.

\subsubsection{Upper Limit Results}
\label{DatabaseDefinitions:ultype}\label{DatabaseDefinitions:experimental-result-upper-limit-type}

\begin{figure}[t!]
	\begin{center}
	\includegraphics[width=0.8\linewidth]{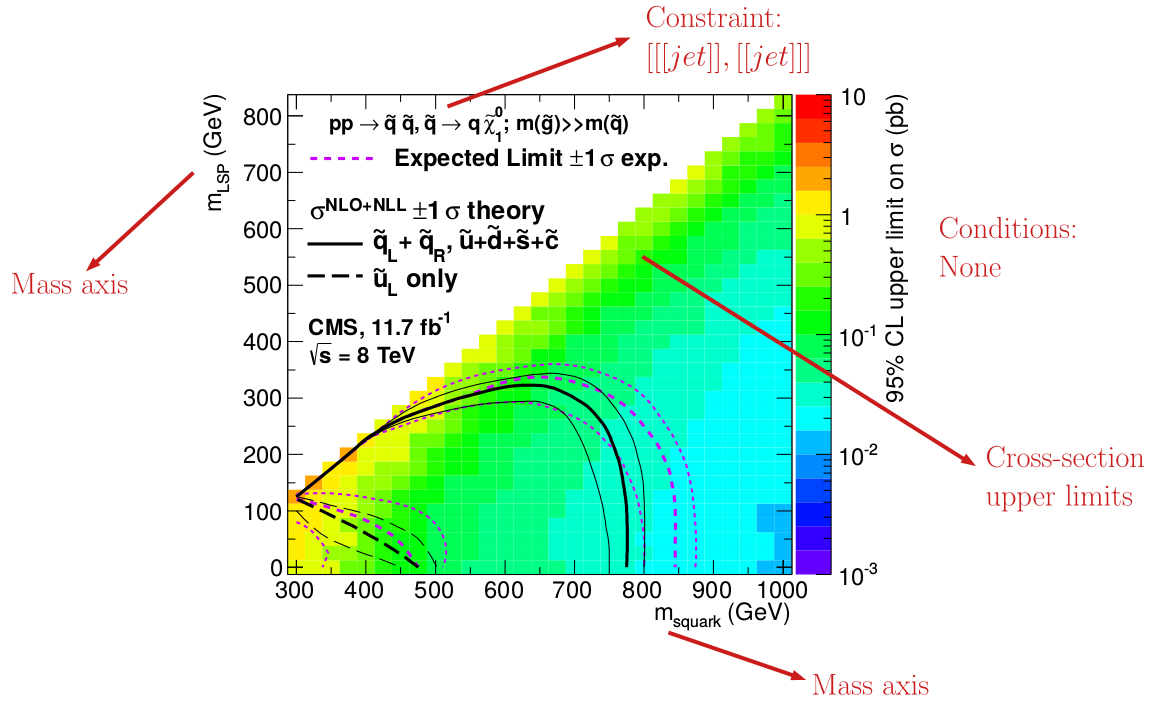}
	\end{center}
\caption{Example of an UL map from~\cite{Chatrchyan:2013lya}. The information used by
SModelS is indicated by arrows.}
\label{DatabaseDefinitions:ulplot}
\end{figure}

Upper Limit experimental results contain the experimental constraints on
the cross section times branching ratio ( \(\sigma \times BR\) ) for
simplified models from a specific experimental publication or preliminary
result. These constraints are typically given in the format of UL maps,
which correspond to 95\% confidence level (CL) upper limit values on \(\sigma
\times BR\) as a function of the respective parameter space (usually BSM masses
or slices over mass planes). The UL values usually assume the best signal region
(for a given point in parameter space), a combination of signal regions or
more involved limits from other methods. An example of an UL map is shown
in Figure~\ref{DatabaseDefinitions:ulplot}.

Within SModelS, the UL map shown in Figure~\ref{DatabaseDefinitions:ulplot} is
used to constrain the simplified model $\tilde{q} + \tilde{q} \to \left(jet+\tilde{\chi}_1^0\right) +
\left(jet+\tilde{\chi}_1^0\right)$.
Using the SModelS notation this simplified model is mapped to the
element \([[[jet]],[[jet]]]\), 
using the notation defined in
Section~\ref{TheoryDefinitions:theory-definitions}.
Usually a single experimental publication contains several UL maps, 
each one constraining different simplified models. 
We stress that the exclusion curve shown in the UL map above is never used by SModelS.

\subparagraph{Upper Limit Constraint}
\label{DatabaseDefinitions:upper-limit-constraint}\label{DatabaseDefinitions:ulconstraint}

\begin{figure}[th!]
	\begin{center}
	\includegraphics[width=0.9\linewidth]{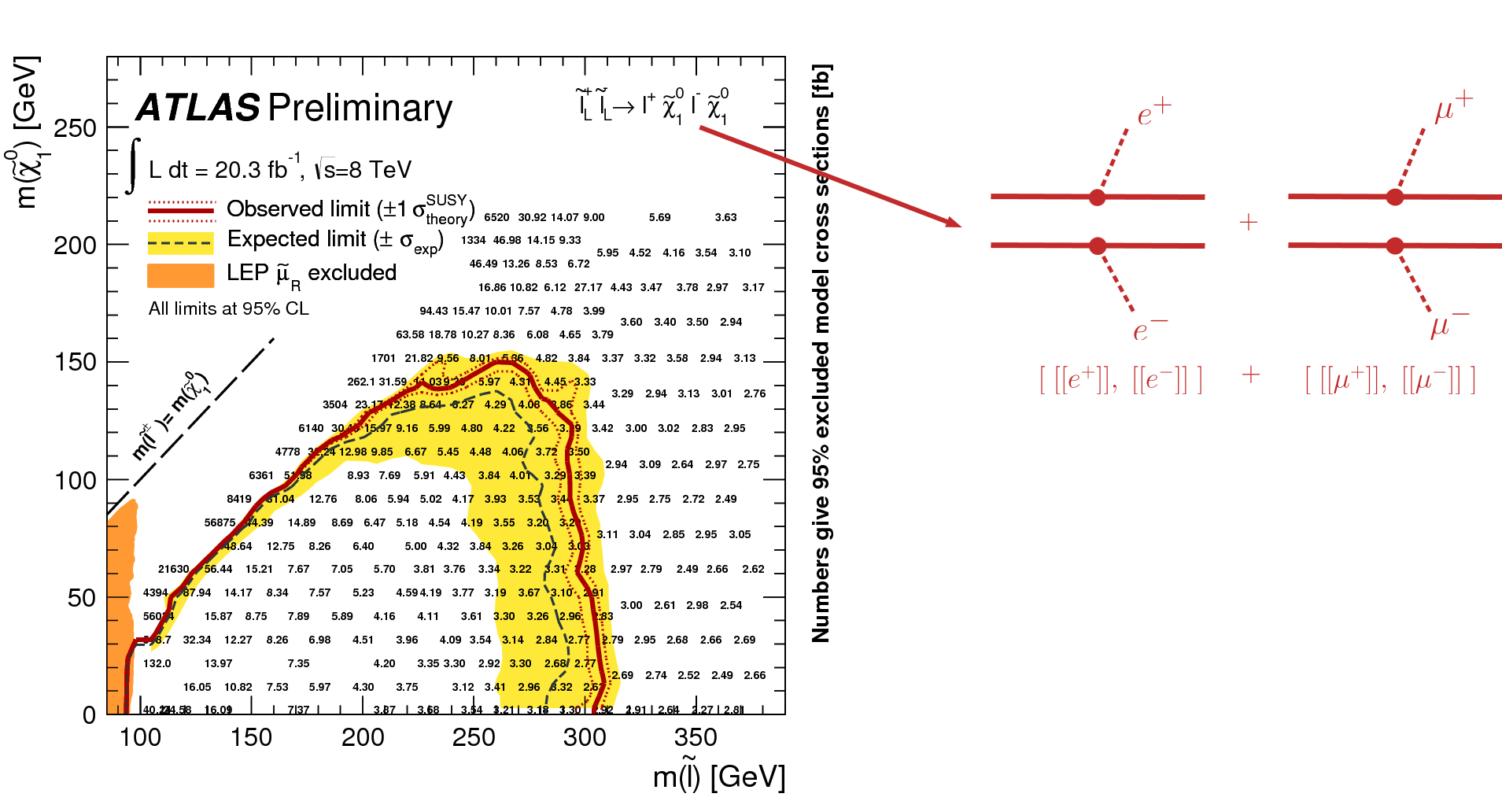}
	\end{center}
\caption{An example of an UL map from \cite{ATLAS-CONF-2013-049} which
constrains  the sum of two elements, correspondingly the UL constraint is
written down as a sum of two elements in the SModelS database.}
\label{DatabaseDefinitions:constraintplot}
\end{figure}

The upper limit constraint specifies which simplified model
is constrained by
the respective UL map. For simple constraints, as the one shown in
Figure~\ref{DatabaseDefinitions:ulplot}, 
it corresponds to a single element (\([[[jet]],[[jet]]]\)). 
In some cases, however, the constraint corresponds to a
sum of elements. As an example,
consider the
ATLAS upper limit map shown 
in Figure~\ref{DatabaseDefinitions:constraintplot}.

Here, the upper limits apply to the sum of the cross sections:
\begin{gather}
	\begin{split}\sigma = \sigma([[[e^+]],[[e^-]]]) + \sigma([[[\mu^+]],[[\mu^-]]])\end{split}\notag
\end{gather}
In this case the UL constraint is:
\begin{gather}
	\begin{split}[[[e^+]],[[e^-]]] + [[[\mu^+]],[[\mu^-]]]\end{split}\notag
\end{gather}
where it is understood that the sum is over the weights of the respective
elements and not over the
elements themselves.

Note that the sum can be over particle charges, flavors or more complex
combinations of elements. However, almost all experimental results sum only over elements
sharing a common topology.

\subparagraph{Upper Limit Conditions}
\label{DatabaseDefinitions:ulconditions}\label{DatabaseDefinitions:upper-limit-conditions}

When the analysis
constraints are
non-trivial (refer to a sum of elements), it is often the case that there are
implicit (or explicit) assumptions about the contribution of each element. For
instance, in Figure~\ref{DatabaseDefinitions:constraintplot},
it is implicitly assumed that each lepton flavor contributes equally to the
summed cross section:
\begin{gather}
	\begin{split}\sigma([[[e^+]],[[e^-]]]) = \sigma([[[\mu^+]],[[\mu^-]]])           \;\;\; \mbox{(condition)}\end{split}\notag
\end{gather}
Therefore, when applying these constraints to general models, one must also
verify that these conditions are satisfied. Once again we can express these
conditions in bracket notation:
\begin{gather}
	\begin{split}[[[e^+]],[[e^-]]] = [[[\mu^+]],[[\mu^-]]]           \;\;\; \mbox{(condition)}\end{split}\notag
\end{gather}
where it is understood that the condition refers to the weights of the
respective elements and not to the elements themselves.

In several cases it is desirable to relax the analysis conditions, so the
analysis upper limits can be applied to a broader spectrum of models. Once
again, for the example mentioned above, it might be reasonable to impose instead:
\begin{gather}
	\begin{split}[[[e^+]],[[e^-]]] \simeq [[[\mu^+]],[[\mu^-]]]           \;\;\; \mbox{(fuzzy condition)}\end{split}\notag
\end{gather}
The \emph{departure} from the exact condition can then be properly quantified
and one can decide whether the analysis upper limits are applicable or not to
the model being considered.
Concretely, SModelS computes for each condition a number between 0 and 1, where
0 means the condition is exactly satisfied and 1 means it is maximally violated.
Allowing for a \(20\%\) violation of a condition corresponds approximately to a
``condition violation value'' (or simply condition value) of 0.2. The condition
values  are given as an output of SModelS, so the user can decide 
on the allowed violation.

\subsubsection{Efficiency Map Results}
\label{DatabaseDefinitions:emtype}\label{DatabaseDefinitions:experimental-result-efficiency-map-type}

Efficiency maps correspond to a grid of simulated acceptance times
efficiency ( \(A \times \epsilon\) ) values for a specific signal region for a
specific simplified model. 
In the following we will refer to \(A \times \epsilon\) simply as \emph{efficiency} and denote it by \(\epsilon\).
Furthermore, additional information, such as the luminosity, number of observed and expected
events, etc is also stored in a EM-type result.

\begin{figure}[th!]
	\begin{center}	
	\includegraphics[width=0.8\linewidth]{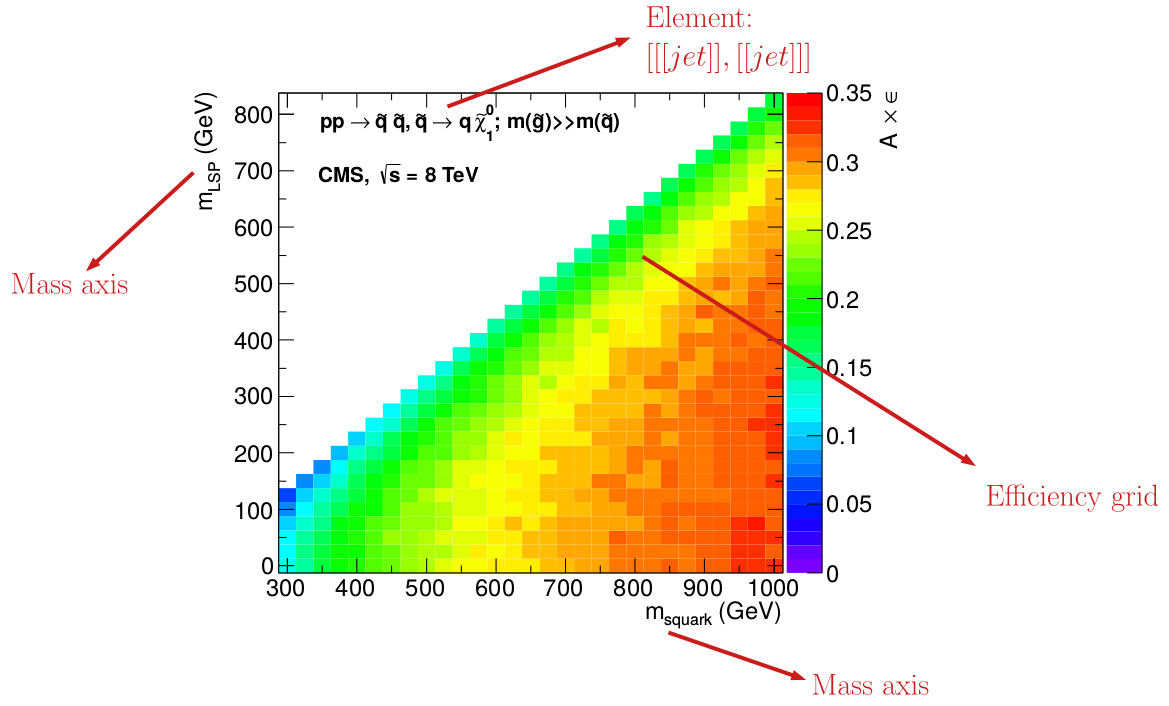}
	\end{center}\vspace*{-4mm}
\caption{An example of an efficiency map from~\cite{Chatrchyan:2014lfa}. The
information used by SModelS is indicated by arrows.}
\label{DatabaseDefinitions:emplot}
\end{figure}

An important difference between UL-type results and EM-type results is the
existence of several signal regions in the latter, which in SModelS are mapped to data sets.
While UL-type results contain a single data set, EM results hold several data
sets, one for each signal region (see
Figure~\ref{DatabaseDefinitions:databasescheme}).
Each data set contains one or more efficiency maps, one for each element or sum
of elements. The efficiency map is usually a function of the BSM masses
appearing in the element, as shown in Figure~\ref{DatabaseDefinitions:emplot}.

The concept of efficiency maps 
can also be extended to UL-type results. For the latter, the efficiencies for a
given element are either 1, if the element appears in the UL constraint, or 0,
otherwise. Atlhough trivial, this extension allows us to treat EM-type results
and UL-type results in a very similar fashion (see
Section~\ref{TheoryPredictions:theorypredictions} for more details).

\subsubsection{TxName Convention}
\label{DatabaseDefinitions:txname-convention}\label{DatabaseDefinitions:txname}

Since using the bracket notation
to describe the simplified models appearing in the upper limit or efficiency
maps can be rather lengthy, it is useful to define a shorthand notation for the
constraints. SModelS
adopts a notation based on the CMS SMS conventions, where each specific
constraint is labeled as
\emph{T\textless{}constraint name\textgreater{}}, which we refer to as
\emph{TxName}. For instance, the TxName corresponding to the constraint in
Figure~\ref{DatabaseDefinitions:constraintplot} is \emph{TSlepSlep}. A complete
list of TxNames can be found in \cite{smodels:dictionary}.

\subsubsection{Database Structure}
\label{DatabaseStructure:experimental-result-folder}

Let us now discuss in more detail how the information about the experimental
results are stored in the database.
As shown in Figure~\ref{DatabaseDefinitions:databasescheme}, the basic
structure is an ordinary (UNIX) directory hierarchy. A thin Python
layer serves as the access to the database.

In the official release, the database is organized according to the
center-of-mass energies (`8TeV' and `13TeV'), followed by  the name of
the collaboration (e.g. `8TeV/ATLAS'). The third level of the
directory hierarchy encodes the experimental results:
\begin{itemize}
    \item {} 8TeV/ATLAS/ATLAS-SUSY-2013-02
    \item {} 8TeV/CMS/CMS-SUS-12-024
    \item {} ...
\end{itemize}
This folder structure is, however, flexible and may be customised by
the user. More as well as fewer subdirectories are allowed, making
for instance a simple division according to the data's provenance
(`LHC', `FastLim', `MadAnalysis5', \ldots) a viable alternative.
One small requirement that is imposed on the database is that the
top-level directory contains a  \code{version} file
with a simple version string. The structure of the experimental result
folders is as follows. Each experimental result folder contains:

\begin{itemize}
    \item {}  a folder for each data set (e.g. \code{data})
    \item {}  a \path{globalInfo.txt} file
\end{itemize}

The \path{globalInfo.txt} file contains the meta information about the
experimental result. It
defines the center-of-mass energy \(\sqrt{s}\), the integrated luminosity, the
id used to identify the result and additional information about the source of
the data. Below we show the content of \path{CMS-SUS-12-024/globalInfo.txt} as an example:

\begin{Verbatim}[commandchars=\\\{\},frame=lines]
sqrts: 8.0*TeV
lumi: 19.4/fb
id: CMS\PYGZhy{}SUS\PYGZhy{}12\PYGZhy{}024
url: https://twiki.cern.ch/twiki/bin/view/CMSPublic/
     PhysicsResultsSUS12024
arxiv: http://arxiv.org/abs/1305.2390
publication: http://www.sciencedirect.com/science/article/pii/
             S0370269313005339
implementedBy: Wolfgang Waltenberger
lastUpdate: 2015/5/11
\end{Verbatim}

\subparagraph{Data Set Folder}
\label{DatabaseStructure:data-set-folder}

Each data set folder 
contains:

\begin{itemize}
    \item {}  the Upper Limit maps for
    UL-type results or Efficiency
    maps for EM-type results (\path{TxName.txt} files)
    \item {}  a \path{dataInfo.txt} file containing meta information about the
    Data Set
\end{itemize}

\subparagraph{Data Set Folder: Upper Limit Type}
\label{DatabaseStructure:data-set-folder-upper-limit-type}

Since UL-type results have a single dataset 
the info file only holds some trivial information, such as the type of
experimental result (UL) and
the dataset id (`None' for UL-type results). For example, the content 
of \path{CMS-SUS-12-024/data/dataInfo.txt} is

\begin{Verbatim}[commandchars=\\\{\},frame=lines]
dataType: upperLimit
dataId: None
\end{Verbatim}

Each
\code{TxName.txt} file contains the UL map for a given simplified model
as well as some meta information, including 
the corresponding constraint and conditions.  The first few lines of
\path{CMS-SUS-12-024/data/T1tttt.txt} read:

\begin{Verbatim}[commandchars=\\\{\},frame=lines]
txName: T1tttt
conditionDescription: None
condition: None
constraint: [[[\PYGZsq{}t\PYGZsq{},\PYGZsq{}t\PYGZsq{}]],[[\PYGZsq{}t\PYGZsq{},\PYGZsq{}t\PYGZsq{}]]]
figureUrl: https://twiki.cern.ch/twiki/pub/CMSPublic/
           PhysicsResultsSUS12024/T1tttt\PYGZus{}exclusions\PYGZus{}corrected.pdf
\end{Verbatim}

The second block contains the upper limits as a function of the BSM masses,
according to the mass array convention defined in Section~\ref{TheoryDefinitions:theory-definitions}. 
It is given as a Python array with the structure 
\code{[[masses,upper limit], [masses,upper limit],...]}. 
Explicitly, continuing the above example:

\begin{Verbatim}[commandchars=\\\{\},frame=lines]
upperLimits: [
[[[400.0*GeV, 0.0*GeV], [400.0*GeV, 0.0*GeV]], 1.815773*pb],
[[[400.0*GeV, 25.0*GeV], [400.0*GeV, 25.0*GeV]], 1.806528*pb],
[[[400.0*GeV, 50.0*GeV], [400.0*GeV, 50.0*GeV]], 2.139336*pb],
[[[400.0*GeV, 75.0*GeV], [400.0*GeV, 75.0*GeV]], 2.472143*pb],
    ...
\end{Verbatim}


\subparagraph{Data Set Folder: Efficiency Map Type}
\label{DatabaseStructure:data-set-folder-efficiency-map-type}

For EM-type results the
\path{dataInfo.txt} contains 
an id to identify the data set (signal region), the
number of observed and expected background events and the error on the number of background events for this signal
region and the respective signal upper limits.  Below we list the content of
\path{CMS-SUS-13-012-eff/3NJet6_1000HT1250_200MHT300/dataInfo.txt} as an example:

\begin{Verbatim}[commandchars=\\\{\},frame=lines]
dataType: efficiencyMap
dataId: 3NJet6\PYGZus{}1000HT1250\PYGZus{}200MHT300
observedN: 335
expectedBG: 305
bgError: 41
upperLimit: 5.681*fb
expectedUpperLimit: 4.585*fb
\end{Verbatim}

Each  \code{TxName.txt} file then contains the efficiency map for a given simplified model
as well as some meta information. Here are the first few lines of
\path{CMS-SUS-13-012-eff/3NJet6_1000HT1250_200MHT300/T2.txt}:

\begin{Verbatim}[commandchars=\\\{\},frame=lines]
txName: T2
conditionDescription: None
condition: None
constraint: [[[\PYGZsq{}jet\PYGZsq{}]],[[\PYGZsq{}jet\PYGZsq{}]]]
figureUrl: https://twiki.cern.ch/twiki/pub/CMSPublic/
           PhysicsResultsSUS13012/Fig\PYGZus{}7a.pdf
\end{Verbatim}

This first block of data in the \code{T2.txt} file contains
information about the element
(\([[[\mbox{jet}]],[[\mbox{jet}]]]\)) in
bracket notation for
 which the efficiencies refers to as well as reference to the original data
 source and some additional information.
The second block of data contains the efficiencies as a function of the BSM
masses,  given as a Python array with the structure
\code{[[masses,efficiency], [masses,efficiency],...]}:

\begin{Verbatim}[commandchars=\\\{\},frame=lines]
efficiencyMap: [
[[[312.5*GeV, 12.5*GeV], [312.5*GeV, 12.5*GeV]], 0.00109],
[[[312.5*GeV, 62.5*GeV], [312.5*GeV, 62.5*GeV]], 0.00118],
[[[312.5*GeV, 112.5*GeV], [312.5*GeV, 112.5*GeV]], 0.00073],
[[[312.5*GeV, 162.5*GeV], [312.5*GeV, 162.5*GeV]], 0.00044],
    ...
\end{Verbatim}


\subsubsection{Binary (Pickle) Format}
\label{DatabaseStructure:database-binary-pickle-format}\label{DatabaseStructure:databasepickle}

At the first time of instantiating the \emph{Database} class, the text files in
\path{database-path} are loaded and parsed, and the corresponding 
data objects are built. The efficiency and upper limit maps themselves are 
subjected to standard preprocessing steps such as a principal component
analysis and Delaunay triangulation.
The simplices defined during triangulation are then used for linearly interpolating the data grid,
thus allowing SModelS to compute efficiencies or upper limits for arbitrary
mass values (as long as they fall inside the data grid).
This procedure provides an efficient and numerically robust way of
dealing with generic data grids, including arbitrary parametrizations of the mass parameter space,
irregular data grids and asymmetric branches.

For the sake of efficiency, the entire database -- including the Delaunay
triangulation -- is then serialized into a pickle
file, which will be read directly the next time the database is loaded.
If any changes in the database folder structure are detected, the Python or the SModelS
version has changed, SModelS will automatically re-build the pickle file. This
action may take a few minutes, but it is performed only once.
If desired, the pickling process can be skipped using the option \emph{force\_load = `txt'}
when running SModelS (see Section~\ref{RunningSModelS:runningsmodels}).

\section{The SModelS Procedure}
\label{Structure:smodels-structure}

In this section we describe in detail the main tasks performed by SModelS:
the simplified model decomposition, the computation of the relevant signal
cross-sections (``theory predictions'') and how these are confronted with the
constraints stored in the database. Finally, we explain how missing topologies are identified. 
First, however, we need to describe how the information about the full model is given as an input.

\subsection{Input Model}
\label{BasicInput:basicinput}\label{BasicInput:index-0}\label{BasicInput::doc}\label{BasicInput:basic-input}

The information about the input (full) model can be given as

\begin{itemize}
	\item {}  an SLHA (SUSY Les Houches Accord~\cite{Skands:2003cj}) file
	containing masses, branching ratios and cross sections for the BSM states, or 

	\item {}  an LHE (Les Houches Event~\cite{Alwall:2006yp}) file containing parton
	level events.
\end{itemize}

The SLHA format is usually more compact and best suited for supersymmetric
models. On the other hand, an LHE file can always be generated for any BSM model
(through the use of your favorite MC generator).\footnote{\smodelsnn\ can easily be used for non-SUSY models as long as they present 
a $\mathbb{Z}_2$-type symmetry. However, as mentioned at the beginning of Section~\ref{SModelSDefs:basic-concepts-and-definitions} 
(see also \cite{Kraml:2013mwa}), a few caveats need to be taken into account. It is the responsibility of the user to make sure that the 
experimental results used actually apply to the model under consideration---if
necessary, a subset of results can be selected via the \code{parameters.ini} file, see Section~\ref{RunningSModelS:the-parameters-file}.
}
In this case, however, the precision of the results is limited to the MC statistics used to generate the file.

\begin{figure}[h!]
	\begin{center}
	\includegraphics[width=1.25\linewidth]{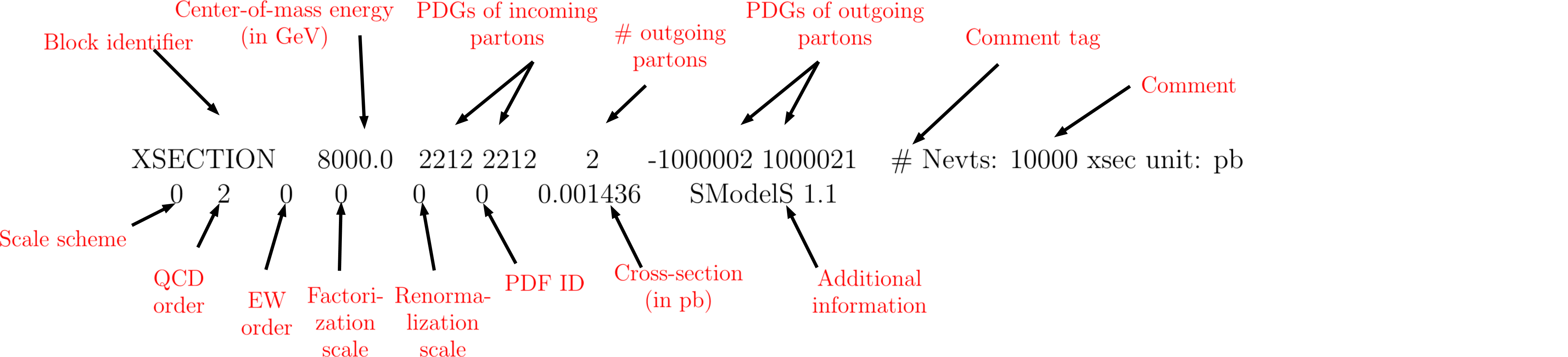}
	\end{center}
\caption{Format of the SLHA cross section block as used in SModelS. Arrows
annotate the meaning of each of the entries in the block.}
\label{BasicInput:xsecblock}
\end{figure}

\emph{\bf In the case of SLHA input}, the production cross sections for the BSM states have to be included in the SLHA file 
following the format defined in~\cite{slha:xsections}.  
Figure~\ref{BasicInput:xsecblock} exemplifies the SLHA cross section format
with \(pp \rightarrow \tilde{u}_L^{\ast} + \tilde{g}\) at a center-of-mass energy of 8 TeV and at NLO+NLL QCD accuracy.
The information used by SModelS are the center-of-mass energy, the outgoing
particle PID's, the cross section value, and the QCD order. \emph{If the input
file contains two cross sections for the same process but at different QCD
orders, only the highest order will be used.}

For the MSSM and some of its extensions,\footnote{Typically extensions by a dark matter candidate other than the neutralino as the lightest SUSY particle (LSP). If the direct production of non-MSSM particles is important, the relevant cross sections have to be computed
by external tools and then added by the user to the SLHA file.} the cross sections may be calculated automatically and 
appended to the SLHA file  using the cross-section calculator tool
provided by SModelS, see Appendix~\ref{Tools:cross-section-calculator}.

\emph{\bf In the case of LHE input}, the total production cross section as well as
the center-of-mass energy should be listed in the
\textless{}init\textgreater{}\textless{}/init\textgreater{} block, according to
the standard LHE format. Moreover, all the
$\mathbb{Z}_2$-even particles (see definition in
Section~\ref{TheoryDefinitions:theory-definitions}) should
be set as stable, since in SModelS they are effectively considered as final
states. When generating the events it is also important to ensure that no mass
smearing is applied, so the mass values for a given particle are the same
throughout the LHE file.

\subparagraph{New Particles}
\label{BasicInput:new-particles}\label{BasicInput:newparticles}

Besides information about the masses and branching ratios, the user must also
define which particles are $\mathbb{Z}_2$-odd states
(intermediate states)
and which are $\mathbb{Z}_2$-even
(final states). These
definitions must be given in the \code{particles.py} file, where some default
values (for SM and MSSM particles) are already loaded.

If the user wants to check the SLHA input file for possible errors (see
Appendix~\ref{Tools:filechecks}), it is also necessary to
define the particle's quantum numbers in the \code{particles.py} file.

\subsection{Decomposition into Simplified Models}
\label{Decomposition:decomposition}\label{Decomposition::doc}\label{Decomposition:decomposition-into-simplified-models}

Given an input model, the first task of SModelS is to decompose it  into a sum of simplified models (or
`elements' in SModelS language).
Based on the input format, SLHA or LHE file, one of two distinct decomposition methods is applied: 
either the SLHA-based or the LHE-based decomposition.

\subsubsection{SLHA-based Decomposition}
\label{Decomposition:slha-based-decomposition}\label{Decomposition:slhadecomp}

The SLHA file describing the input model is required to contain the masses of
all the BSM states as well as their production cross sections and decay
branching ratios. All of this information must follow the SLHA file standard.

\begin{figure}[h!]
	\begin{center}
	\includegraphics[width=0.80\linewidth]{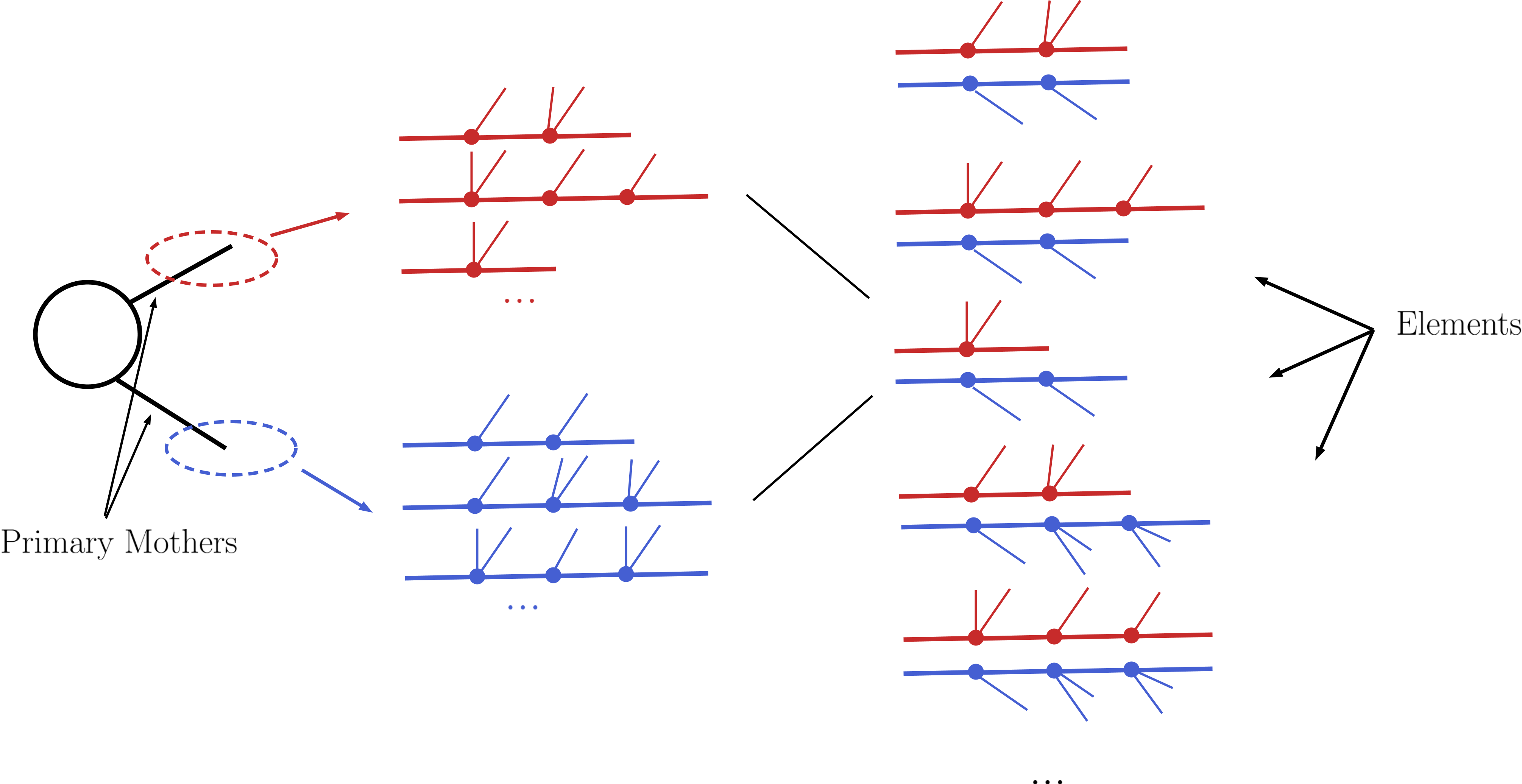}
	\end{center}
\caption{SModelS decomposition process: the diagram represents the creation of
elements given the primary mother pair production and subsequent branches.}
\label{Decomposition:decomp2}
\end{figure}

Once the production cross sections are read from the input file, all the
cross sections for \emph{production of two} $\mathbb{Z}_2$-odd \emph{states} are
stored and serve as the initial step for the decomposition. (All the other
cross sections with a different number of $\mathbb{Z}_2$-odd states are
ignored.) Starting from these primary mothers, all the possible decays are
generated according to the information contained in the DECAY blocks. This
procedure is represented by the first decomposition step shown in
Figure~\ref{Decomposition:decomp2}.
Each of the possible cascade decays for each mother corresponds to a
branch. In order to finally
generate elements, all the
branches are combined in pairs according to the branching ratios,
as shown by the second step in Figure~\ref{Decomposition:decomp2}.

Each of the elements generated
according to the procedure just described will also store its weight, which
equals the element production cross section times all the branching ratios
appearing in it. In order to avoid a too large number of elements, only those satisfying a
{\it minimum weight requirement} are kept. Furthermore, the elements are grouped
according to their topologies. The final output of
the SLHA decomposition is a list of such topologies, where each topology
contains a list of the elements generated during the decomposition.

\subparagraph{Minimum Decomposition Weight}
\label{Decomposition:minweight}\label{Decomposition:minimum-decomposition-weight}

Some models may contain a large number of new states and each may have a large
number of possible decays. As a result, long cascade decays are possible and the
number of elements generated by the decomposition process may become too large,
resulting in excessively long computing times. For most practical purposes,
however, elements with very small weights 
can be discarded, since
they will fall well below the experimental limits. Therefore, during the SLHA
decomposition, whenever an element is generated with a weight below some minimum
value, this element, and all elements which would be derived from it, are ignored. 
The minimum weight to be considered is set by the {\tt sigmacut} parameter and is easily
adjustable, see Section~\ref{RunningSModelS:parameterfile}.

Note that, when computing the theory predictions, the weight of several elements
can be combined together. Hence it is recommended to set the value of
{\tt sigmacut} approximately one order of magnitude below the minimum signal
cross sections the experimental data can constrain.

\subsubsection{LHE-based Decomposition}
\label{Decomposition:lhe-based-decomposition}\label{Decomposition:lhedecomp}

\begin{figure}[h!]\centering
\includegraphics[width=\linewidth]{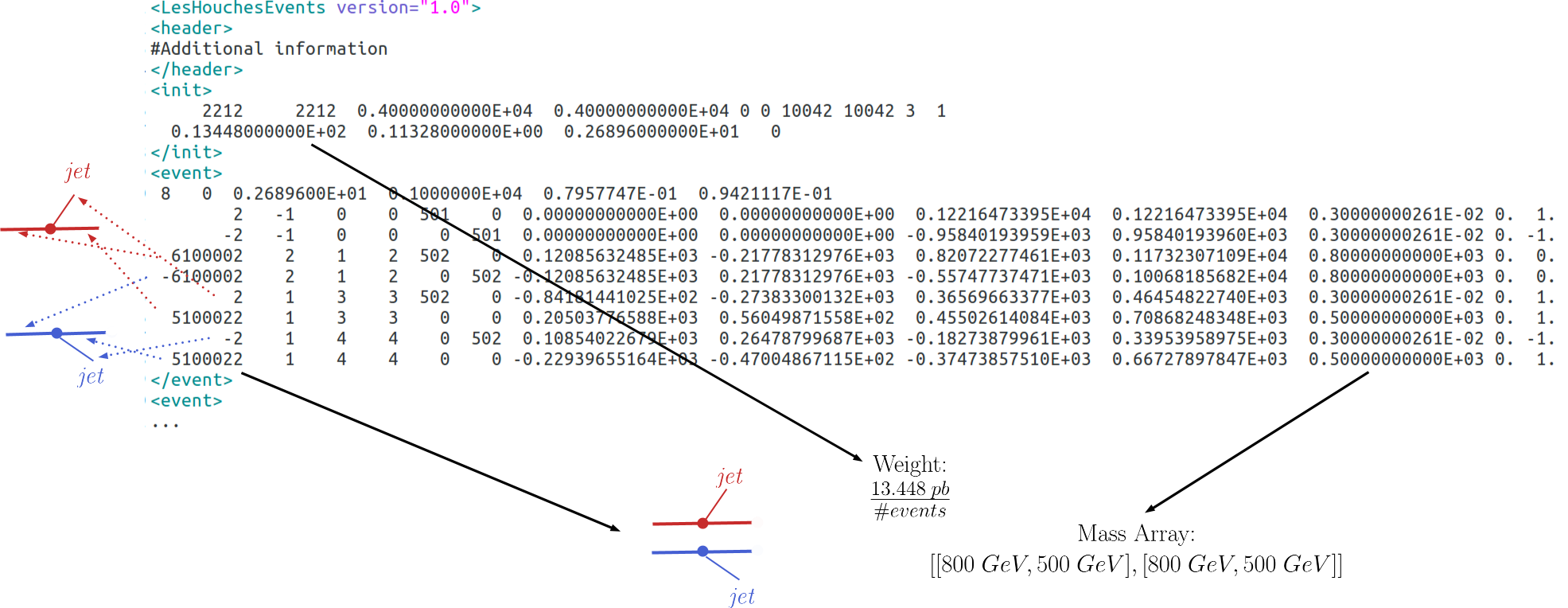}
\caption{An example of an LHE event. Arrows indicate
the information used by SModelS to construct the corresponding element.}
\label{Decomposition:event}
\end{figure}

More general models can be input through an LHE event file containing
parton-level events, including the production of the primary mothers and their
cascade decays. Each event can then be directly mapped to an
element with the element weight
corresponding to the event weight. Finally, identical elements can be combined
together (adding their weights). How the information from an LHE event
is used to construct a SModelS element is
illustrated in Figure~\ref{Decomposition:event}.

Notice that, for the LHE decomposition, the
elements generated are restricted
to the events in the input file. Hence, the uncertainties on the elements
weights (and which elements are actually generated by the model) are fully
dependent on the Monte Carlo statistics used to generate the LHE file. Also,
when generating the events it is important to ensure that no mass smearing is
applied, so the events always contain the same mass value for a given particle.

\subsubsection{Compression of Elements}
\label{Decomposition:compression-of-elements}\label{Decomposition:elementcomp}

During the decomposition process it is possible to perform several
simplifications on the elements generated. In both the LHE and SLHA-based
decompositions, two useful simplifications are possible:
mass compression and invisible compression. The main advantage of performing
these compressions is that the simplified element is always shorter (has fewer
cascade decay steps), which makes it more likely to be constrained by
experimental results. The details behind the compression methods are as follows:

\subparagraph{Mass Compression}
\label{Decomposition:masscomp}\label{Decomposition:mass-compression}

\begin{figure}[b!]
	\begin{center}
	\includegraphics[width=0.90\linewidth]{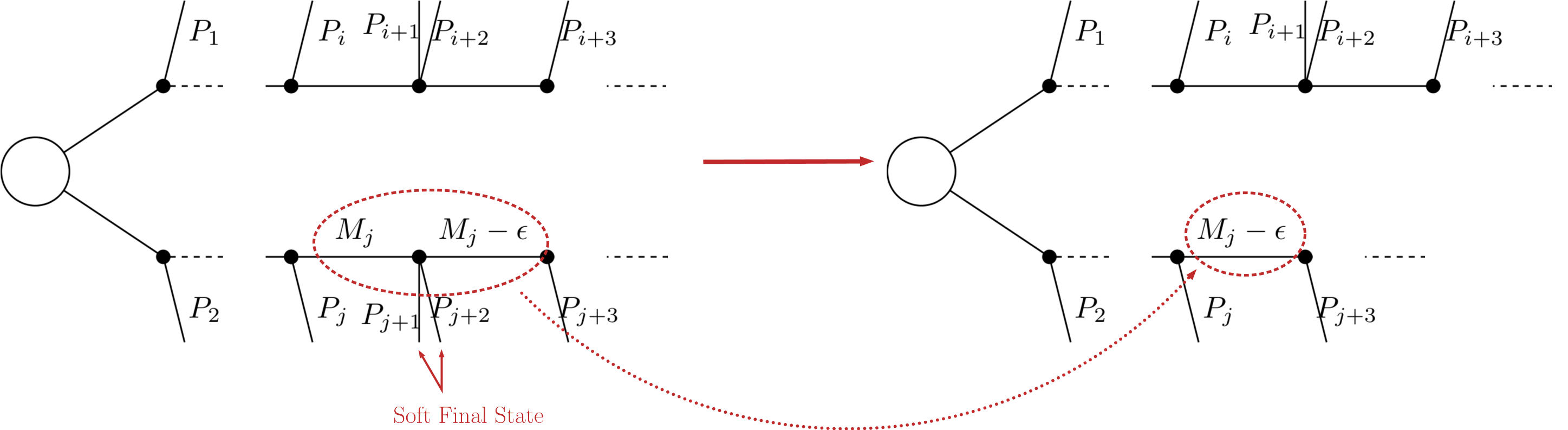}
	\end{center}
\caption{Schematic representation of `mass compression' performed by SModelS to
deal with soft final states.}
\label{Decomposition:masscompfig}
\end{figure}

In case of small mass differences, the decay of an
intermediate state to a
nearly degenerate one will in most cases produce soft
final states, which
can not be experimentally detected. Consequently, it is a good approximation to
neglect the soft final states 
and \emph{compress} the respective decay, as shown in Figure~\ref{Decomposition:masscompfig}. 

After the compression, only the lightest of the two near-degenerate masses are
kept in the element, as shown in Figure~\ref{Decomposition:masscompfig}.
The main parameter which controls the compression is called \code{minmassgap} (see
Section~\ref{RunningSModelS:parameterfile}),
which corresponds to the maximum value of \(\epsilon\) in Figure~\ref{Decomposition:masscompfig}
up to which the compression is performed:

\begin{gather}
\begin{split}& \mbox{if } |M_j - M_{j+1}| < {\tt minmassgap} \rightarrow \mbox{the decay is compressed}\\
& \mbox{if } |M_j - M_{j+1}| > {\tt minmassgap} \rightarrow \mbox{the decay is NOT compressed}\\\end{split}\notag
\end{gather}

Note that the compression is an approximation since the final states, depending on the boost of the parent state, may not always be soft. It is recommended to choose values of \code{minmassgap} of 1--10~GeV; the default value is 5 GeV.

\subparagraph{Invisible Compression}
\label{Decomposition:invisible-compression}\label{Decomposition:invcomp}

\begin{figure}[h!]
	\begin{center}
	\includegraphics[width=0.900\linewidth]{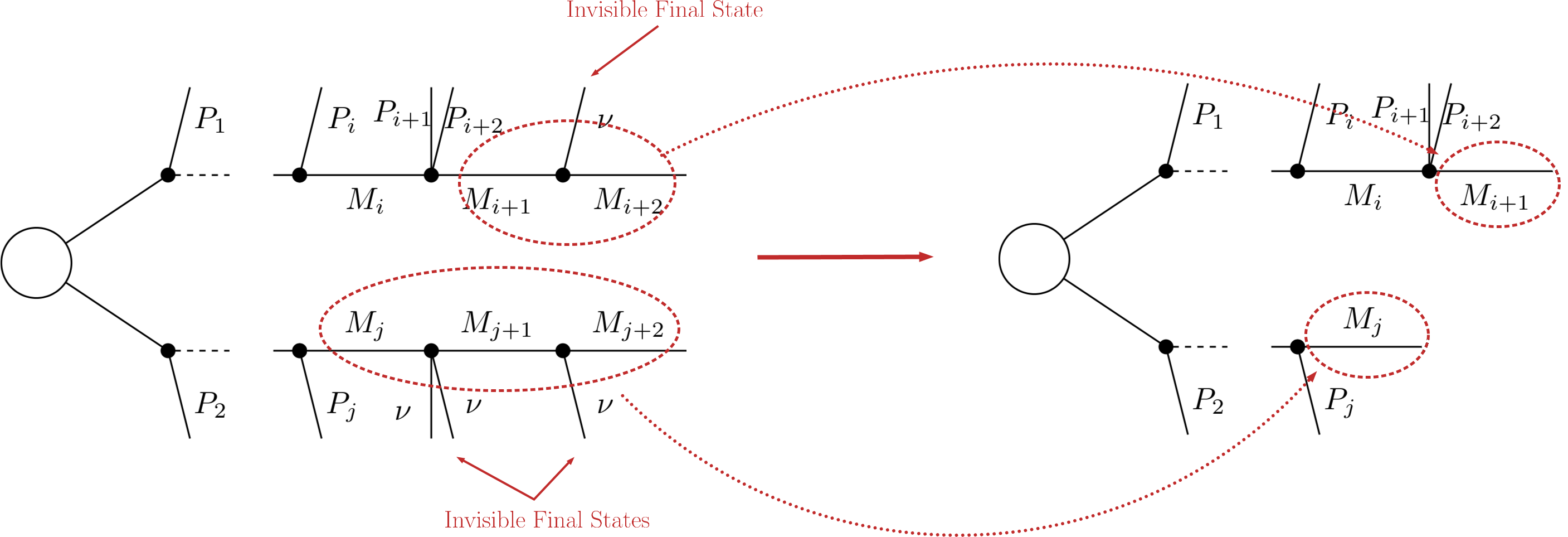}
	\end{center}
\caption{A schematic representation of `invisible compression' as performed by
SModelS to deal with the emission of invisible particles in the final steps of
the cascade decay.}
\label{Decomposition:massinvpfig}
\end{figure}

Another type of compression is possible when the
final states of the
last decay 
are invisible. 
The most common example is
\begin{gather}
	\begin{split}A \rightarrow \nu + B\end{split}\notag
\end{gather}
as the last step of the decay chain, where \(B\) is an invisible particle
leading to a MET signature. Since both the neutrino and \(B\) are invisible, for
all experimental purposes the effective MET object is \(B + \nu = A\). Hence it
is possible to omit the last step in the cascade decay, resulting in a
compressed element. Note that this compression can be applied consecutively to
several steps of the cascade decay if all of them contain only invisible final
states, as illustrated in Figure~\ref{Decomposition:massinvpfig}.

\subsection{Computing Theory Predictions}
\label{TheoryPredictions:theory-predictions}\label{TheoryPredictions:theorypredictions}\label{TheoryPredictions::doc}

After the decomposition of the input model into simplified models, the next
step consists of computing the relevant signal cross sections (or \emph{theory
predictions}) for comparison with the experimental limits.

Note that UL and EM-type results each require different theoretical predictions
to be compared against experimental limits. While UL-type results constrain the
total weight (\(\sigma \times BR\)) of a given simplified model, EM-type results
constrain the total signal cross-section (\(\sum \sigma
\times BR \times \epsilon\)) in a given signal region (data set).
We deal with both types of results in an equal footing by defining 
\begin{equation}
    \mbox{theory prediction } = \sum_{elements} (\sigma \times BR
    \times \epsilon) 
\end{equation}
where $\epsilon$ is a ``generalized efficiency'': for EM-type results it
corresponds to the simplified model signal efficiency, while for UL-type results
it equals to 1 (0) if the element is (is not) constrained by the experimental
result.
With this definition the theory prediction value can then be directly
compared to the respective 95\%~CL upper limit on the signal
cross-section extracted from the UL map (for UL-type results) or 
with the signal upper limit for the respective signal region (for EM-type
results).

\begin{figure}[h!]
	\begin{center}
	\includegraphics[width=\linewidth]{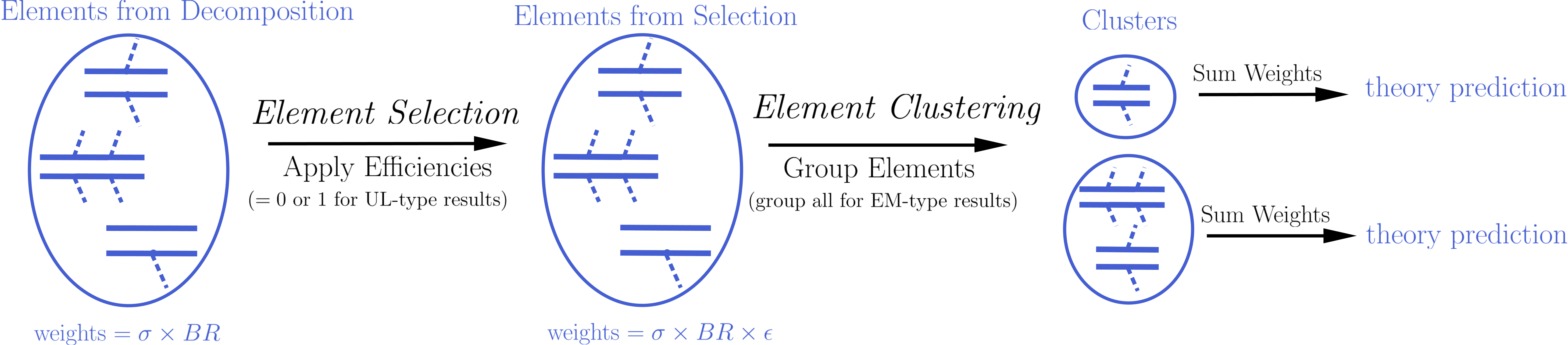}
	\end{center}
\caption{SModelS procedure for computing the
signal cross sections. The generated
elements undergo selection and clustering procedure before the final theory
prediction can be compared with the corresponding experimental limit.}
\label{TheoryPredictions:theoryPredScheme}
\end{figure}

The calculation of the theory prediction can always be divided in two main steps:
\emph{Element Selection} and \emph{Element Clustering}, as shown schematically
in Figure~\ref{TheoryPredictions:theoryPredScheme}.
These steps are described in more detail below.

\subsubsection{Element Selection}

Given a specific experimental result (either EM-type or UL-type),
the first step for computing the theory predictions
is to select the simplified models which are constrained by the corresponding
result.
For EM-type results this corresponds to selecting all elements which
contain non-zero efficiencies ($\epsilon$) for a given signal region, while for
UL-type results this step implies selecting all elements which are constrained
by the corresponding UL map (see Section~\ref{DatabaseDefinitions:database-definitions}).

During this step, the selected elements have their weights 
(\(\sigma \times BR\)) rescaled by their corresponding efficiencies.
As mentioned above, for UL-type results these efficiencies are trivial and
equal 1 (0) if the element appears (does not appear) in the UL constraint.
On the other hand, for EM-type results, these efficiencies are obtained from the
efficiency maps stored in the database for the corresponding experimental
result/data set.

\begin{figure}[t!]
	\begin{center}
	\includegraphics[width=0.950\linewidth]{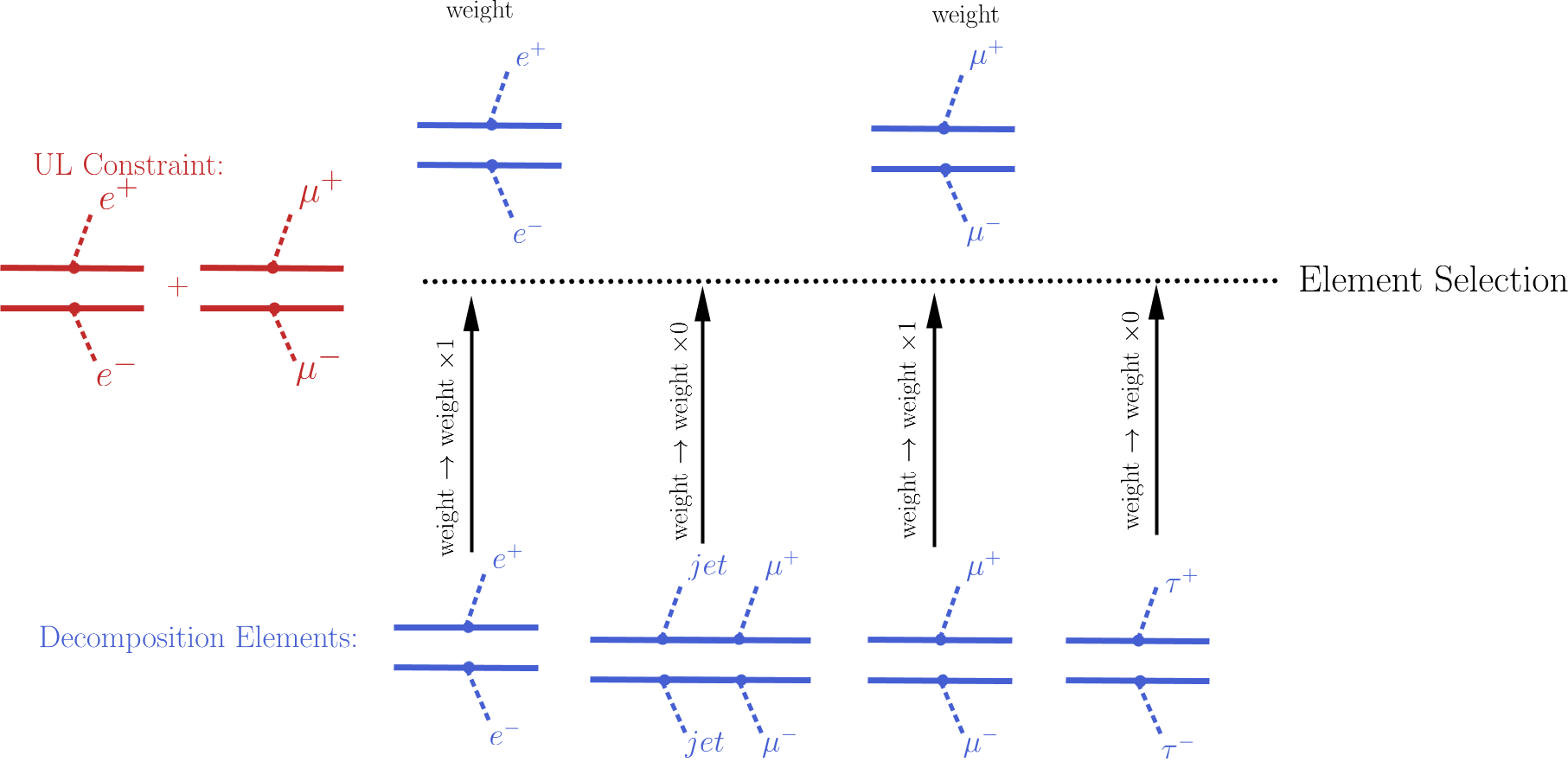}
	\end{center}
\caption{Procedure for selecting elements for an UL-type result.}
\label{TheoryPredictions:ULselection}
\end{figure}

\begin{figure}[t!]
\label{TheoryPredictions:emselectionfig}
    \begin{center}
    \includegraphics[width=0.950\linewidth]{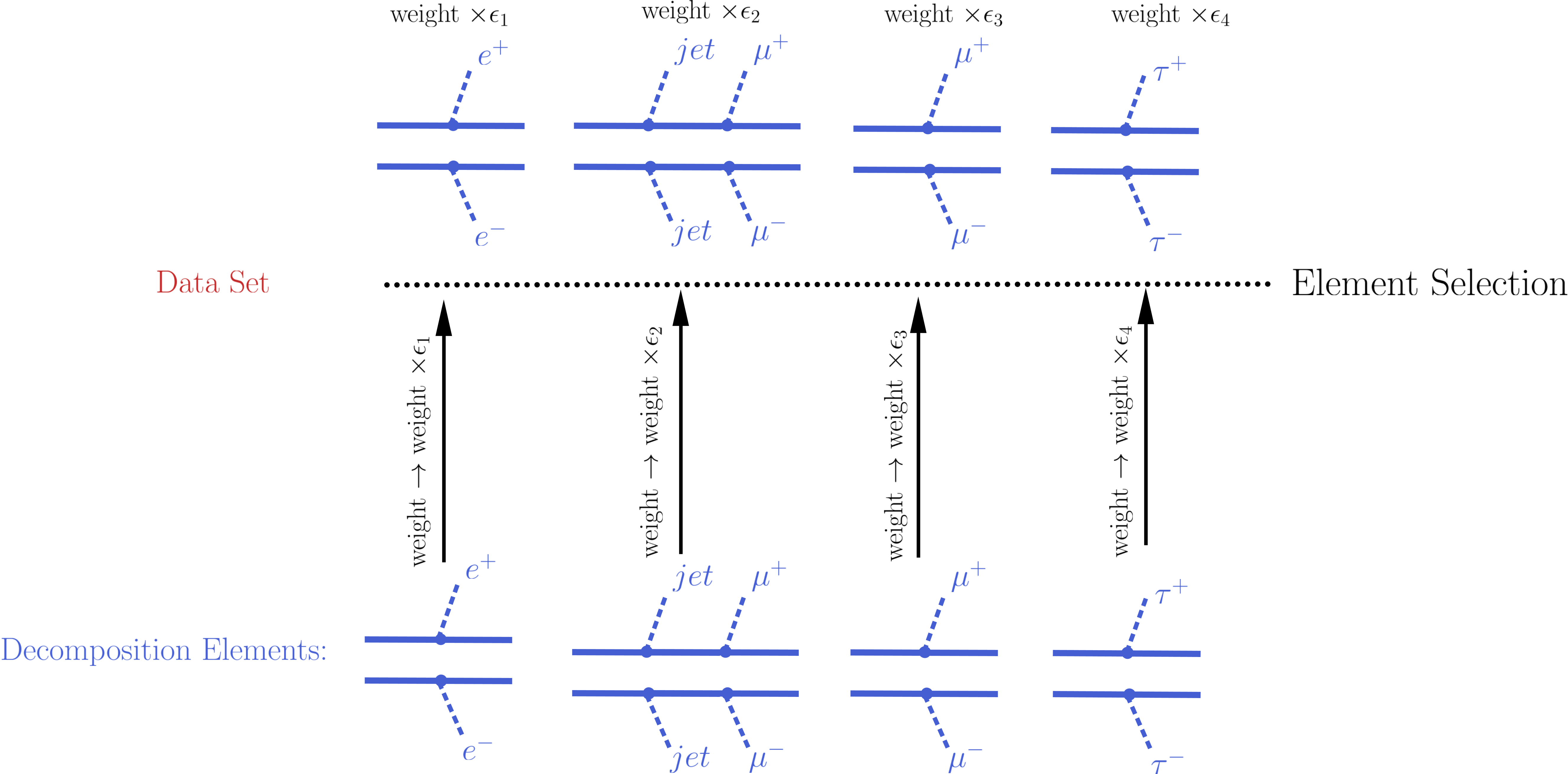}
    \end{center}
\caption{Procedure for selecting elements for EM-type results.}
\label{TheoryPredictions:EMselection}
\end{figure}

At the end of the element selection step only the elements with non-zero
(rescaled) weights are relevant for computing the theory prediction for the corresponding
experimental result and all the other elements are ignored.
The procedure just described is illustrated graphically in
Figures~\ref{TheoryPredictions:ULselection} and
~\ref{TheoryPredictions:EMselection} for an UL-type and EM-type result,
respectively.

\subsubsection{Element Clustering}
\label{TheoryPredictions:element-clustering}\label{TheoryPredictions:ulcluster}

Naively one would expect that after all the
elements constrained by the experimental result have been selected and
their weights have been rescaled by the corresponding efficiencies, it is
trivial to compute the theory prediction. One must simply sum up the weights
(\(\sigma \times BR \times \epsilon\)) of all the selected elements. 
This is indeed the case for EM-type results, where the final
theory prediction is simply given by the sum of the selected element weights.

For UL-type results, however, the experimental
limit on $\sigma \times BR$ only applies to elements with the same
mass\footnote{When referring to an element mass, we mean all the intermediate state masses
appearing in the element (see Section~\ref{TheoryDefinitions:theory-definitions}
for more details). Two elements are considered to have identical masses if their
mass arrays are identical .} (or mass array). As a result, the selected
elements must be grouped into \emph{clusters} of equal masses. When grouping the
elements, one can allow for small mass differences, since the
experimental efficiencies should not be strongly sensitive to them.
For instance, assume two elements contain identical mass arrays,
except for the parent masses which differ by 1 MeV. In this case it is obvious
that for all experimental purposes the two elements have identical masses and
should contribute to the same theory prediction (\eg, their weights should be
added when computing the signal cross section).
Unfortunately there is no way to unambiguously define ``similar masses'' and the
definition should depend on the experimental result, since different results
will be more or less sensitive to mass differences. SModelS uses an UL
map-dependent measure of the distance between two element masses, as described
below.

\begin{figure}[th!]
	\begin{center}
	\includegraphics[width=0.900\linewidth]{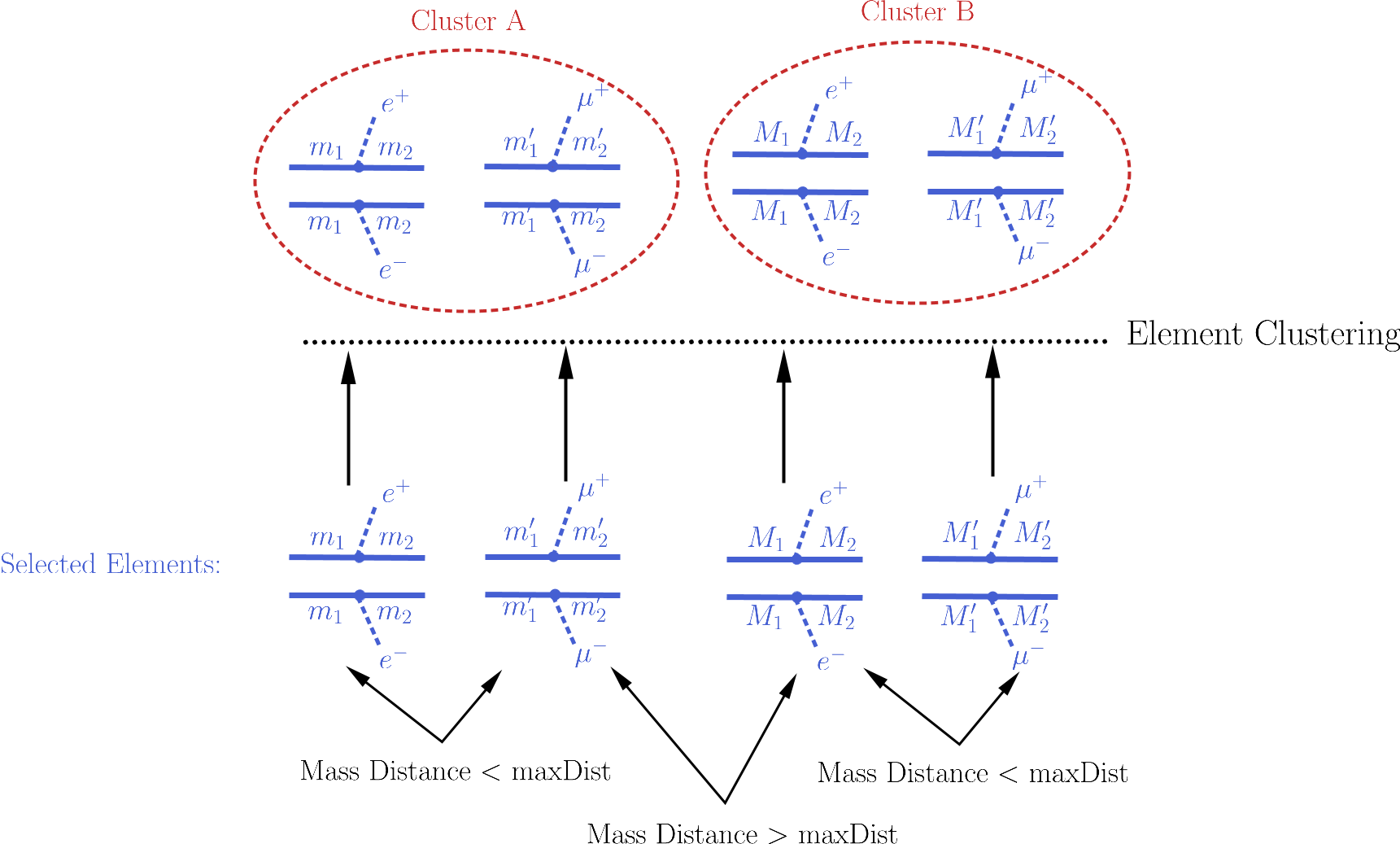}
	\end{center}
\caption{Element clustering procedure in SModelS.}
\label{TheoryPredictions:ULcluster}
\end{figure}

If two of the selected elements have a mass distance smaller than a maximum
value (defined by {\tt maxDist}), they are grouped in the same mass cluster, as
illustrated in Figure~\ref{TheoryPredictions:ULcluster}.
Once all the elements have been
clustered, their weights can finally be added together and compared against the
experimental upper limit.

\subparagraph{Mass Distance}
\label{TheoryPredictions:massdist}\label{TheoryPredictions:mass-distance}

As mentioned above, for UL-type results it is necessary to group
elements with similar masses. Since an absolute definition of
``similar masses'' is not possible and the sensitivity to mass differences
depends on the experimental result, SModelS uses an ``upper limit
map-dependent'' definition. For each element`s mass array, the upper limit for
the corresponding mass values is obtained from the UL map. 
This way, each mass array is mapped to a single number (the cross section upper
limit for the experimental result). Then the distance between the two
elements'
masses is simply given by the relative difference between their respective upper
limits. More explicitly:
\begin{align}
  \mbox{Element } A\; (& M_A = [[M1,M2,...],[m1,m2,...]]) \rightarrow \mbox{ Upper Limit}(M_A) = x \notag\\
  \mbox{Element } B\; (& M_B = [[M1',M2',...],[m1',m2',...]]) \rightarrow \mbox{ Upper Limit}(M_B) = y \notag\\
                                    & \Rightarrow \mbox{mass distance}(A,B) = \frac{2|x-y|}{(x+y)} 
\end{align}
where \(M_A,M_B\) (\(x,y\)) are the mass arrays (upper limits) for the
elements A and B, respectively.
If the mass distance of two
elements is smaller than {\tt maxDist},
the two masses are considered similar. The default value
for {\tt maxDist} in SModelS is 0.2.

Notice that the above definition of mass distance quantifies the experimental
analysis' sensitivity to mass differences, which is the relevant parameter when
clustering elements. Also, a
check is performed to ensure that masses with very distinct values but similar
upper limits are not clustered together.

\subsection{Confronting Predictions with Experimental Limits}
\label{ConfrontPredictions:confrontpredictions}\label{ConfrontPredictions::doc}
\label{ConfrontPredictions:confronting-predictions-with-experimental-limits}

\subparagraph{R-Values}
Once the relevant signal cross sections (theory predictions)
have been computed for the input model, they must be compared to the respective
experimental constraints.  SModelS reports the result in the form of $r$-values
defined as
\begin{equation}
  r = {\textrm{(theory prediction)/(upper limit)}}.
\end{equation}

In the case of an UL-type result,
the theory predictions typically consist of a list of signal cross sections (one
for each cluster). 
Each theory prediction must then be compared to its corresponding upper limit. 
This limit is simply the cross section upper limit provided by the experimental
publication or preliminary result and is extracted from the corresponding UL map
(see Section~\ref{DatabaseDefinitions:database-definitions}).

For EM-type results there is a
single theory prediction ($\sum \sigma\times BR\times \epsilon$) for each data set (or
signal region). This value must be compared to the upper limit on the number of
signal events for the corresponding signal region. This upper limit is easily
computed using the number of observed and expected events for the data set and their
uncertainties and is typically stored in the
database.
Since most EM-type results have several signal regions (data sets), there will
be one theory prediction/upper limit for each data set. By default SModelS
considers only the best data set, \ie\ the one with the largest value for 
$r_{exp} = \mbox{(theory prediction)}/\mbox{(expected limit)}$.
In this case each EM-type result will have a single $r$-value, corresponding  to
the signal region with the best expected limit.

The procedure described above can be applied to all the experimental results in
the database, resulting in a list of theory predictions and upper limits for
each experimental result. A model can then be considered excluded by the
experimental results if, for one or more predictions, we have $r>1$.%
\footnote{The statistical significance of the exclusion statement is difficult
to quantify exactly, since the model is being tested by a large number of
results simultaneously.}
SModelS can also identify the relevant simplified model topologies which have large
cross-sections, but are not constrained by any of the experimental results
in the database, see Section~\ref{Tools:topology-coverage}.

\subparagraph{Likelihood Computation}
\label{ConfrontPredictions:likelihoodcalc}\label{ConfrontPredictions:likelihood-computation}
\newcommand{\LL}{\mathcal{L}}

In the case of EM-type results, additional statistical information
about the constrained model can be provided by the SModelS output.
Following Ref.~\cite{simplifiedlikelihoods}, we construct a simplified
likelihood which describes the plausibility of the data $D$, given a signal strength $\mu$:

\begin{equation}
	\LL(\mu,\theta|D) =  P\left(D|\mu + b + \theta \right) p(\theta) \,.
\end{equation}

\noindent
Here, $\theta$ denotes the nuisance parameter that describes the
variations in the signal and background contributions due to systematic
effects. We assume $p(\theta)$ to follow a Gaussian distribution centered
around zero and with a variance of $\delta^2$,
whereas $P(D)$ corresponds to a counting variable and is thus
properly described by a Poissonian. The complete likelihood thus reads

\begin{equation}
	\LL(\mu,\theta|D) = \frac{(\mu + b + \theta)^{n_{obs}} e^{-(\mu + b +
\theta)}}{n_{obs}!} exp \left( -\frac{\theta^2}{2\delta^2} \right) \,,
\end{equation}

\noindent
where $n_{obs}$ is the number of observed events in the signal region.
A test statistic $T$ can now be constructed from a likelihood ratio test:

\begin{equation}
   T = -2 \ln \frac{H_0}{H_1} = -2 \ln \left(\frac{\LL(\mu=n_{\mathrm{signal}},\theta|D)}{sup\{\LL(\mu,\theta|D) : \mu \in \mathbb{R}^+ \}}\right) \,.
\end{equation}

As the signal hypothesis in the numerator presents a special case of the
likelihood in the denominator, the Neyman-Pearson lemma holds, and we
can assume $T$ to be asymptotically distributed according to a $\chi^2$ distribution
with one degree of freedom.
Of course care must taken in deep Poissonian cases
where both $b$ and $\theta$ are small, as the test statistic in this case
cannot be expected to follow a $\chi^2$ distribution anymore.
Because $H_0$ assumes the signal strength of
a particular model, $T=0$ corresponds to a perfect match between that
model's prediction and the measured data. $T \gtrsim 3.84$ corresponds to
a 95\% CL upper limit (in the asymptotic case). 

While \(n_{\mathrm{obs}}\), $b$ and \(\delta_{b}\) are directly extracted from
the data set
(coined {\tt observedN}, {\tt expectedBG} and {\tt bgError}, respectively),
\(n_{\mathrm{signal}}\) is obtained from the calculation of the
theory predictions. A default 20\% systematical uncertainty is assumed for \(n_{\mathrm{signal}}\),
resulting in $\delta^2 = \delta_{b}^2 + \left(0.2 n_{\mathrm{signal}}\right)^2$.

\smodelsnn\ reports the $\chi^2$ ($T$ values) and likelihood {\it for each} EM-type result,
together with the observed and expected $r$ values.
We note that in the general case analyses may be correlated, so summing up the $T$ values will no longer follow a $\chi^2_{(n)}$ distribution.
Therefore, for a conservative interpretation, only the result with the best expected limit should be used.
Moreover, for a statistically rigorous usage in scans, it is recommended to check that the analysis giving the
best expected limit does not wildly jump within continuous regions of parameter space that give roughly the same phenomenology.


\subsection{Identifying Missing Topologies}
\label{Tools:topcoverage}\label{Tools:topology-coverage}

The constraints provided by SModelS are obviously limited by 
the available set of simplified model interpretations in the database.
Therefore it is interesting to identify classes of missing simplified models
(termed `missing topologies') which are relevant for a given input model, but are
not constrained by the results in the SModelS database. This task is performed
as a last step in SModelS, once the decomposition and the theory predictions
have been computed.

Using the decomposition output, the elements (see
Section~\ref{TheoryDefinitions:theory-definitions}) which are not tested by any
of the experimental results in the database are grouped into the following classes:%
\footnote{If mass or invisible compression are turned on, elements which can be compressed are not considered, to avoid double counting.} 

\begin{itemize}
    \item {} 
    \emph{missingTopos}: elements whose final
    states are not tested by any of the experimental results in the database
    (independent of the element mass). The missing topologies are further
    classified as:
    \begin{itemize}
            \item {} 
            \emph{longCascade}: elements with long cascade decays (more than one
            intermediate particle in one of the branches);
            
            \item {} 
            \emph{asymmetricBranches}:
            elements with one branch differing from the other (not
            considering cases of long cascade decays, see above).
    \end{itemize}      

    \item {} 
    \emph{outsideGrid}: elements which could be tested by one or more
    experimental results, but are not constrained because the mass array falls
    outside the experimental mass grids.
\end{itemize}

Usually the list of  \emph{missingTopos} or \emph{outsideGrid} elements is of 
considerable length. Hence, to compress this list, all elements differing only by
their masses (with the same final states) or electric charges are combined
into a \emph{missing} or \emph{outsideGrid} topology.
Moreover, by default, electrons and muons are combined to light leptons (denoted
``l''); gluons and light quarks are combined into jets.
The \emph{missing} topologies are then further classified (if applicable) into
\emph{longCascade} or \emph{asymmetricBranches} topologies.

The topologies for each of the four categories are then grouped according to the
final state (for the \emph{missingTopos} and \emph{outsideGrid} classes) or
according to the PID's of the initially produced mother particles (for the
\emph{longCascade} and \emph{asymmetricBranches} classes).
For the latter, elements deriving from different mother particles but with the
same final state and mass array cannot be distinguished and are therefore
combined. The full list of mother PID's is provided in the Python output or as
a comment in both the `stdout' and `summary' outputs.

The topology coverage tool is normally called from within SModelS (\eg, when
running \code{runSModelS.py}) by setting \textbf{testCoverage=True} in the parameters
file (see Section~\ref{RunningSModelS:parameterfile}). In the output,
contributions in each category are ordered by cross section. 
By default only the ten with the largest cross sections are shown.

\index{Using SModelS}
\section{Using SModelS}
\label{RunningSModelS:running-smodels}\label{RunningSModelS:runningsmodels}\label{RunningSModelS::doc}

\smodelsnn\  ships with a command-line tool, \code{runSModelS.py}, 
which is deemed configurable enough to cover a large range of applications.
The functionalities contained in this tool include detailed checks of input SLHA or LHE files, 
running the
decomposition, evaluating the
theory predictions and
comparing them to the experimental limits available in the
database, determining
missing topologies and printing the
output in several available formats.
Moreover, starting from v1.1.0, \code{runSModelS.py} can process a whole folder containing a set of SLHA or LHE
files in a parallelized fashion. 

The command-line tool and the parameter file are described in detail below.
Users familiar with Python and the SModelS basics may however prefer to write their own main routine. 
A simple example code is provided in \code{Example.py} and explained step-by-step in the html manual.

Finally, \smodelsnn\ can also conveniently be used within micrOMEGAS, as explained
in \cite{Barducci:2016pcb}.


\subsection{Usage of runSModelS.py}
\label{RunningSModelS:runsmodels-py}\label{RunningSModelS:runsmodels}

\begin{Verbatim}[commandchars=\\\{\},frame=lines]
runSModelS.py {[}-h{]} -f FILENAME {[}-p PARAMETERFILE{]} {[}-o OUTPUTDIR{]}
              {[}-d{]} {[}-t{]} {[}-V{]} {[}-c{]} {[}-v VERBOSE{]} {[}-T TIMEOUT{]}
\end{Verbatim}

\begin{description}
\item[{\emph{arguments}:}] \leavevmode
\begin{optionlist}{3cm}
\item [-h, -{-}help]  show this help message and exit.
\item [-f FILENAME, -{-}filename FILENAME]  name of SLHA or LHE input file or a
directory path (required argument). If a directory is given, loop over all files in the directory.
\item [-p PARAMETERFILE, -{-}parameterFile PARAMETERFILE] name of the parameter file, where most options are defined (optional argument). If not set, use all parameters from
{\code{smodels/etc/\allowbreak parameters\_default.ini}}.
\item [-o OUTPUTDIR, -{-}outputDir OUTPUTDIR] name of output directory (optional
argument). The default folder is {\tt ./results/}. 
\item [-d, -{-}development]  if set, SModelS will run in development mode and exit if any errors are found 
\item [-t, -{-}force\_txt] force loading the text database.
\item [-V, -{-}version]  show program's version number and exit
\item [-c, -{-}run-crashreport] parse crash report file and use its contents for
a SModelS run. Supply the crash file simply via `-- filename myfile.crash'.
\item [-v VERBOSE, -{-}verbose VERBOSE]  sets the verbosity level (debug, info, warning, error). Default value is ``info''.
\item [-T TIMEOUT, -{-}timeout TIMEOUT]  define a limit on the running time (in
secs). If not set, run without a time limit. If a directory is given as input, the timeout will be
  applied for each individual file.
\end{optionlist}
\end{description}

\noindent
A typical usage example is:

\begin{Verbatim}[commandchars=\\\{\}]
runSModelS.py \PYGZhy{}f inputFiles/slha/gluino_squarks.slha \\
              \PYGZhy{}p parameters.ini \PYGZhy{}o ./ \PYGZhy{}v warning
\end{Verbatim}

The resulting output will be
generated in the current folder, according to the printer options set in the
parameter file.

\subsection{Parameter File}
\label{RunningSModelS:the-parameters-file}\label{RunningSModelS:parameterfile}

The basic options and parameters used by \code{runSModelS.py} are defined in the
parameter file. An example parameter file, including all available parameters
together with a short description, is stored in \code{parameters.ini}. If no
parameter file is specified, the default parameters stored in
\code{smodels/etc/\allowbreak parameters\_default.ini} are used. Below we describe each entry in the parameter file.

\noindent\begin{description}
\item[\emph{options}:] main options for turning SModelS features on or off
\end{description}

\begin{itemize}
  
  \item {}
  \textbf{checkInput} (True/False): if True, \code{runSModelS.py} will run the
  file check tool on the input file and
  verify that the input contains all the necessary information.
  
  \item {}
  \textbf{doInvisible} (True/False): turns
  invisible compression on or off
  during the decomposition. 
  
  \item {}
  \textbf{doCompress} (True/False): turns
  mass compression on or off during
  the decomposition. 
  
  \item {}
  \textbf{computeStatistics} (True/False): turns the likelihood and \(\chi^2\)
  computation on or off for EM-type results (see
  Section~\ref{ConfrontPredictions:confronting-predictions-with-experimental-limits}).
  
  \item {}
  \textbf{testCoverage} (True/False): set to True to run the
  coverage tool (see Section~\ref{Tools:topology-coverage}).
\end{itemize}

\noindent\begin{description}
\item[\emph{parameters}:] basic parameter values for running SModelS
\end{description}

\begin{itemize}
  \item {}
  \textbf{sigmacut} (float): minimum value for an
  element weight (in fb).  Elements with a weight below \code{sigmacut} are neglected during
  decomposition of SLHA input files (see Section~\ref{Decomposition:slhadecomp}).
  The default value is $0.03$~fb. Note that, depending on the input model,
  the running time may increase considerably if
  sigmacut is too low, while too large values might eliminate relevant elements.
  
  \item {}
  \textbf{minmassgap} (float): maximum value of the mass difference (in GeV) for
  perfoming mass compression.
  The default value is $5$~GeV. 
  \emph{Only used if} \code{doCompress = True}.
  
  \item {}
  \textbf{maxcond} (float): maximum allowed value (in the {[}0,1{]} interval)
  for the violation of upper limit conditions. A zero value means the conditions are strictly enforced,
  while 1 means the conditions are never enforced.
  \emph{Only relevant for printing the} output summary.
  
  \item {}
  \textbf{ncpus} (int): number of CPUs. When processing multiple SLHA/LHE files,
  SModelS can be run in a parallelized fashion, splitting up the input files in
  equal chunks. \code{ncpus=-1} uses the total number of CPU cores of the
  machine.
    
\end{itemize}

\noindent\begin{description}
\item[\emph{database}:] allows for selection of a subset of
experimental results from the
database
\end{description}

\begin{itemize}

  \item {}
  \textbf{path}: the absolute or relative path to the
  database. The user can
  supply either the directory name of the database, or the path to the
  pickle file (see Section~\ref{DatabaseStructure:databasepickle}).

  \item {}
  \textbf{analyses} (list of results): set to {\emph{all}} to use all available results.
  If a list of experimental analyses is given, only these will be used. For instance, setting
  \code{analyses = ATLAS-SUSY-2015-09, CMS-PAS-SUS-15-002} will only use the
  experimental results from these two analyses.
    
  \item {}
  \textbf{txnames} (list of topologies): set to {\emph{all}} to use all available
  simplified model topologies.  The topologies are labeled
  according to the TxName convention.  If a list of TxNames are
  given, only the corresponding topologies will be considered.
  For instance, setting \code{txnames=T2} will only consider experimental results
  for $pp \to \tilde{q} + \tilde{q} \to (jet+\tilde{\chi}_1^0) + (jet+\tilde{\chi}_1^0)$ and the output will only contain
  constraints for this topology. A list of all topologies and their corresponding TxNames can be found at~\cite{smodels:dictionary}.
  
  \item {}
  \textbf{dataselector} (list of datasets): set to {\emph{all}} to use all available
  data sets. If dataselector = upperLimit (efficiencyMap), only UL-type results
  (EM-type results) will be used. Furthermore, if a list of signal regions
  (data sets) is given, only the experimental results
  containing these datasets will be used. For instance, if 
	\code{dataselector = SRA mCT150, SRA mCT200}, 
	only these signal regions will be used.

  \item {}
  \textbf{discardZeroes} (True/False): set to True to discard all efficiency maps with zero-only entries.
  
\end{itemize}

\noindent\begin{description}
\item[\emph{printer}:] main options for the
output format
\end{description}

\begin{itemize}
  \item {}
  \textbf{outputType} (list of outputs): use to list all the output formats to
  be generated. Available output formats are: summary, stdout, log, python, xml and slha.
\end{itemize}

\noindent\begin{description}
\item[\emph{stdout-printer}:] options for the stdout or log printer
\end{description}

\begin{itemize}
  \item {}
  \textbf{printDatabase} (True/False): set to True to print the list of selected
  experimental results to
  stdout. 
  
  \item {}
  \textbf{addAnaInfo}  (True/False): set to True to include detailed information
  about the TxNames tested by
  each experimental result.
  \emph{Only used if printDatabase=True}.
  
  \item {}
  \textbf{printDecomp} (True/False): set to True to print basic information from
  the decomposition
  (topologies, total weights, ...).
  
  \item {}
  \textbf{addElementInfo}  (True/False): set to True to include detailed
  information about the elements
  generated by the decomposition. 
	\emph{Only used if printDecomp=True}.
  
  \item {}
  \textbf{printExtendedResults} (True/False): set to True to print extended
  information about the  theory predictions, including the PID's of the particles contributing to the
  predicted cross section, their masses and the expected upper limit (if available).
  
  \item {}
  \textbf{addCoverageID} (True/False): set to True to print the list of element
  IDs contributing to each missing topology (see
  coverage). \emph{Only used if
  testCoverage = True}. This option should be used along with
  \emph{addElementInfo = True} so the user can precisely identify which elements
  were classified as missing.
\end{itemize}

\noindent\begin{description}
\item[\emph{summary-printer}:] options for the summary printer
\end{description}

\begin{itemize}
  \item {}
  \textbf{expandedSummary} (True/False): set to True to include in the summary
  output all applicable
  experimental results, False
  for only the strongest one.
\end{itemize}

\noindent\begin{description}
\item[\emph{python-printer}:] options for the Python printer
\end{description}

\begin{itemize}
  \item {}
  \textbf{addElementList} (True/False): set to True to include in the Python output
  all information about all
  elements generated in the
  decomposition. If set to True
  the output file can be quite large.
\end{itemize}

\noindent\begin{description}
\item[\emph{xml-printer}:] options for the xml printer
\end{description}

\begin{itemize}
  \item {}
  \textbf{addElementList} (True/False): set to True to include in the xml output
  all information about all
  elements generated in the
  decomposition. If set to True
  the output file can be quite large.
\end{itemize}

\subsection{Output}
\label{RunningSModelS:the-output}\label{RunningSModelS:smodelsoutput}

The results of \code{runSModelS.py} are printed in the format(s) specified by the
\textbf{outputType} in the parameter file.
The following formats are available:
\begin{itemize}
  \item {}
  a human-readable screen output (stdout) or log output.
  These are intended to provide detailed information about the
  database, the  decomposition, the  theory predictions and the
  missing topologies. The complexity of the output can be controlled through several options in the
  parameter file. Due to its size, this output is not suitable for storing the results from a large scan; 
  it is more appropriate for a single file input.

  \item {}
  a human-readable text file output containing a summary of the run.
  This format contains the main SModelS results: the theory predictions and the
  missing topologies, as described in detail below. It can be used for
  a large scan, since the output can be made quite compact, using the options in
  the parameter file.
  
  \item {}
  a Python dictionary printed to a file containing information about the
  decomposition, the theory predictions, and the
  missing topologies. The output can be significantly long, if all options in the
  parameter file are set to
  True. However this output can be easily imported to a Python environment,
  making it easy to access the desired information. For users familiar with
  the Python language this is the recommended format.
  
  \item {}
  an xml file containing
  information about the
  decomposition, the
  theory predictions and the
  missing topologies. The output can be
  significantly long, if all options are set to True. 

  \item {}
  a .smodelsslha file (outputType=slha) containing a summary of the most
constraining results, and the missing topologies in the SLHA-type
  format described in~\cite{Barducci:2016pcb}.
\end{itemize}

\vspace*{6mm}

\hrule
\vspace*{2mm}
\noindent {\bf Notes:}
\begin{itemize}
  \item {}
  The list of elements can be 
  extremely long. Try setting \textbf{addElementInfo} = False and/or
  \textbf{printDecomp} = False to obtain a smaller output. 
  \item{}
  A comment of caution is in order regarding a potentially naive use of the highest $r$-value
  reported by SModelS, as this does not necessarily come from the most sensitive
  analysis. For a rigorous statistical interpretation, one should use the
  $r$-value of the result with the highest \emph{expected} $r$ ($r_{exp}$).
  Unfortunately, for UL-type results, the expected limits are often not
  available; $r_{exp}$ is then reported as N/A in the SModelS output.
  \item{}
  We also point out that the $r$-values do not include systematic uncertainties for the signal prediction. The
  signal uncertainties may be relevant, in particular in cases where cross sections are computed only at LO.\\
\hrule
\end{itemize}


\bigskip

As an example we explain below the text-type (summary) output obtained from the
sample file \path{gluino_squarks.slha} in \path{inputFiles/slha/}.
The output file is written in terms of the following blocks:

\begin{itemize}\item information about the basic input parameters and the status of the run:\end{itemize}

\begin{Verbatim}[frame=none,fontsize=\footnotesize]
Input status: 1
Decomposition output status: 1 #decomposition was successful
# Input File: inputFiles/slha/gluino_squarks.slha
# maxcond = 0.2
# minmassgap = 5.
# ncpus = 1
# sigmacut = 0.03
# Database version: 1.1.1
\end{Verbatim} 

\begin{itemize}
  \item a list of all the theory predictions obtained and the corresponding
  experimental result upper limit. If \code{expandedSummary = False} only the most
  constraining experimental result is printed. For each applicable experimental
  result, the corresponding experimental result ID and the center-of-mass energy
  (sqrts) are  printed together with the theory cross section (`Theory\_Value'),
  the observed upper limit (`Exp\_limit'), the (theory cross section)/(observed
  upper limit) ratio (r) and, when available, the expected $r$ value
  (r\_expected). Moreover, the condition violation is given for UL-type results;
  for EM-type results the signal region used is printed. Finally, the TxNames
  contributing to the signal cross section are listed and, if \code{computeStatistics
  = True}, the $\chi^2$ and likelihood values are printed for EM-type results:\end{itemize}

\begin{Verbatim}[fontsize=\footnotesize]
#Analysis Sqrts Cond_Violation Theory_Value(fb) Exp_limit(fb) r r_expected
 
 CMS-SUS-13-019  8.00E+00    0.0  1.773E+00  3.760E+00  4.716E-01  N/A
 Signal Region:  (UL)
 Txnames:  T2
--------------------------------------------------------------------------------
 ATLAS-SUSY-2013-02  8.00E+00    0.0  6.617E+00  1.718E+01  3.851E-01  N/A
 Signal Region:  (UL)
 Txnames:  T6WW
--------------------------------------------------------------------------------
 ATLAS-SUSY-2013-02  8.00E+00    0.0  5.525E-01  1.818E+00  3.039E-01  3.653E-01
 Signal Region:  SR2jt
 Txnames:  T1, T2
 Chi2, Likelihood =  4.185E-02  2.542E-02
--------------------------------------------------------------------------------
 ...
\end{Verbatim} 

\begin{itemize}\item the maximum value for the (theory cross section)/(observed
upper limit) ratio. If this value is higher than 1, the input model is likely
excluded by one of the experimental results.
\end{itemize}

\begin{Verbatim}[frame=none,fontsize=\footnotesize]
The highest r value is = 0.471627309932
\end{Verbatim}

\begin{itemize}\item summary information about the missing topologies, if
\code{testCoverage = True}. The total `missing topology' cross section corresponds to
the sum of weights of all elements which are not tested by any experimental
result. If an element is constrained by one or more experimental results, but
its mass is outside the efficiency or upper limit grids, its cross section is
instead included in the ``outside the mass grid'' category. Finally, the
elements which contribute to the total missing topology cross section are
subdivided into elements with long decays (cascades with more than one
intermediate particle) and with asymmetric branches.
\end{itemize}

\begin{Verbatim}[frame=none,fontsize=\footnotesize]
Total missing topology cross section (fb):  2.767E+02
Total cross section where we are outside the mass grid (fb):  1.760E-01
Total cross section in long cascade decays (fb):  1.096E+02
Total cross section in decays with asymmetric branches (fb):  1.630E+02
\end{Verbatim}

\begin{itemize}\item detailed information about the missing topologies, sorted
by their cross sections. The element cross section (weight) as well as its
description in bracket notation is included. For definiteness, the entries in
this and the following lists are sorted by their weights. 
\end{itemize}

\begin{Verbatim}[frame=none,fontsize=\footnotesize]
Missing topologies with the highest cross sections (up to 10):
Sqrts (TeV)   Weight (fb)        Element description
  8.0  1.601E+01    #                 [[[jet],[W]],[[jet,jet],[W]]]
  8.0  1.395E+01    #       [[[jet],[jet,jet],[W]],[[jet,jet],[W]]]
  8.0  9.206E+00    #                 [[[b,t],[W]],[[jet,jet],[W]]]
  ...
\end{Verbatim}

\begin{itemize}\item detailed information about the topologies which are outside
the experimental results data grid:
\end{itemize}

\begin{Verbatim}[frame=none,fontsize=\footnotesize]
Contributions outside the mass grid (up to 10):
Sqrts (TeV)   Weight (fb)        Element description
  8.0  1.440E-01    #                             [[[jet]],[[t,t]]]
  8.0  3.203E-02    #                         [[[t],[W]],[[t],[W]]]
\end{Verbatim}

\begin{itemize}\item information about the missing topologies with long cascade
decays. The long cascade decays are classified by the initially produced mother
particles. If more than one pair of mothers contribute to the same class
of elements, the full list is given in the comment. 
\end{itemize}

\begin{Verbatim}[frame=none,fontsize=\footnotesize]
Missing topos: long cascade decays (up to 10 entries), sqrts = 8 TeV:
Mother1 Mother2 Weight (fb) # allMothers
1000021 2000002  3.743E+01 # [[1000021, 2000002]]
1000002 1000021  1.626E+01 # [[1000002, 1000021]]
1000021 2000001  1.282E+01 # [[1000021, 2000001]]
 ...
\end{Verbatim}

\begin{itemize}\item information about the missing topologies with asymmetric decays, in the same format as the long cascade decay description:
\end{itemize}

\begin{Verbatim}[frame=none,fontsize=\footnotesize]
Missing topos: asymmetric branches (w/o long cascades, up to 10), sqrts = 8 TeV
Mother1 Mother2 Weight (fb) # allMothers
1000002 1000021  4.725E+01 # [[1000002, 1000021]]
1000021 1000021  4.324E+01 # [[1000021, 1000021]]
1000021 2000002  2.215E+01 # [[1000021, 2000002]]
 ...
\end{Verbatim}

\section{Conclusions and Outlook}\label{Conclusions}

SModelS is an automatised tool for interpreting
simplified model results from the LHC. It can decompose the signatures of any
BSM model containing a $\mathbb{Z}_2$ symmetry into its SMS topologies and
compare them to the existing LHC constraints from a large database of
experimental results.

Version~1.1 of the code, presented in this paper, includes several new
features, most importantly the use of efficiency maps. Efficiency
maps allow us to combine the results from different topologies
and thus improve the constraining power of the tool.
\smodels\ is also equipped with a likelihood and $\chi^2$ calculator useful for a basic statistical
interpretation of the results. Moreover, extended information is provided on the
topology coverage. Several speed improvements are further included in order
to allow for a faster testing of BSM models.

Not counting results marked as `superseded by a more recent analysis',
the database v1.1.1 is comprised of 186 results (125 upper limits and 61 efficiency maps) 
from 21 ATLAS and 23 CMS SUSY searches~\cite{smodels:listofanalyses}, covering a total 
of 37 simplified models. 
From these 44 searches, 11 (4 from ATLAS and 7 from CMS) are based on early 13~TeV Run~2 data, 
comprising 35 UL-type and 2 EM-type results. 
FastLim-1.0 efficiency maps converted to \smodels\ 
format are also available; they cover another 15 simplified models. 

A new database browser tool is provided for easy access to the information
stored in the database.  An update of the database with more 13~TeV results is
in preparation and will be released soon, as will be new `home-grown'
efficiency maps for testing topologies currently absent in the \smodelsnn\
database.

\begin{figure}[t!]\centering
\includegraphics[width=0.8\linewidth]{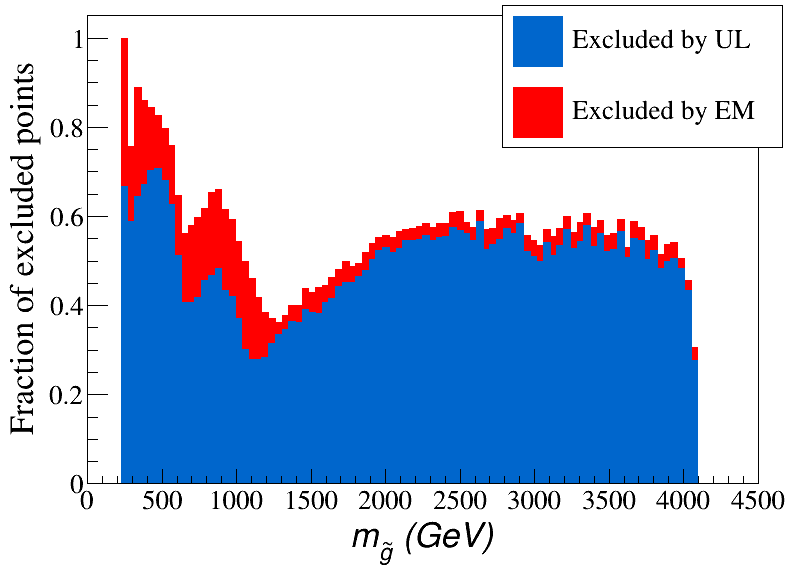}
\caption{Fraction of the ATLAS-excluded pMSSM points~\cite{Aad:2015baa}
also excluded by \smodelsnn\ as a function of the gluino mass. The exclusion using only UL-type results (blue) is
contrasted to the exclusion achieved by using also EM-type results (red). Note that this is based on 8 TeV results only; 
points with long-lived sparticles are not considered.}
\label{pMSSMgluinoPlot}
\end{figure}

To illustrate the importance of the extension to efficiency maps presented here, we have tested
how \smodels\ constrains the publicly available points from the ATLAS pMSSM study~\cite{Aad:2015baa}. 
For a direct comparison with ATLAS, we here consider only 8~TeV results in the database. 
Points with long-lived BSM particles, which currently cannot be treated in \smodelsnn, are discarded. 
The fraction of points excluded by ATLAS which is also excluded by \smodelsnn\
is shown in Figure~\ref{pMSSMgluinoPlot} as a function of the gluino mass. 
As can be seen, the gain from using efficiency maps is quite substantial, in
particular for gluinos within LHC reach that exhibit a variety of different decay modes.%
\footnote{A detailed study of the coverage of the pMSSM by simplified model results will be presented elsewhere.}

Generally, the improvement achieved by using efficiency maps is 
particularly relevant for regions of parameter space where the signal
cross section is spread over several competing simplified model
topologies. 
We therefore ask the ATLAS and CMS collaborations to provide as much as possible efficiency maps for each (aggregated) signal region in addition to the UL maps---this will considerably enhance the usefulness of their SMS interpretations.
The alternative is 
to produce `home-grown' efficiency maps based on implementations of the experimental analyses in 
public recast frameworks like \MA~\cite{Conte:2014zja,Dumont:2014tja} or \CM~\cite{Drees:2013wra,Dercks:2016npn}  
as presented in Appendix~\ref{EMcreation}. 
This is, however, clearly less precise than using `official' efficiency maps.

In conclusion, \smodels\ marks an important milestone for the SModelS collaboration, and
towards the efforts of using SMS results in a systematic fashion. However, the
mission is far from complete: extending the software to non-MET final states
is but one out of several major improvements envisaged for future versions of
SModelS.

\section*{Acknowledgements}

We thank the ATLAS and CMS SUSY groups for helpful discussions on their results,
and in particular for providing SMS cross section upper limits and efficiency
maps in digital format.

This work was supported in part by the French ANR, project DMAstro-LHC
ANR-12-BS05-0006, and the Theory-LHC-France Initiative of the CNRS (INP/IN2P3).
F.A.\ is supported by the Austrian FWF, project P26896-N27,
Su.K.\ by the ``New Frontiers'' program of the Austrian Academy of
Sciences, U.L.\ by  the ``Investissements d'avenir, Labex ENIGMASS'', and
A.L.\  by the S\~ao Paulo Research Foundation (FAPESP), projects 2015/20570-1 and 2016/50338-6.
The work of J.S.\ is supported by the collaborative research center
SFB676  ``Particles, Strings, and the Early Universe'' by the German
Science Foundation (DFG) and by the German Federal
Ministry of Education and Research (BMBF).



\renewcommand{\thesection}{\Alph{section}}
\setcounter{section}{0}
\renewcommand*{\theHsection}{chX.\the\value{section}}

\section{Installation and Deployment}\label{Installation}


\subsection{Standard Installation}
\label{Installation:standard-installation}

SModelS is Python code that requires Python version 2.6 or later with the
Python packages \code{setuptools, unum, numpy, argparse, docutils} ($\ge
0.3$), \code{scipy} ($\ge 0.9.0$), and \code{pyslha} ($\ge 3.1.0$).

The cross section computer provided by \code{smodelsTools.py}
makes use of \pythiaSF~\cite{Sjostrand:2006za} or
\pythiaET~\cite{Sjostrand:2006za,Sjostrand:2014zea}
together with \nllfast 1.2 (7 TeV), 2.1 (8 TeV), and 3.1 (13 TeV)
\cite{Beenakker:1996ch,Beenakker:1997ut,Kulesza:2008jb,Kulesza:2009kq,Beenakker:2009ha,Beenakker:2010nq,Beenakker:2011fu}.
These programs need not be downloaded separately: \pythiaSFTS and \nllfast are included in the \smodels\ distribution, 
while \pythiaE will be downloaded and compiled automatically when required.
A fortran compiler (preferably gfortran) is needed to compile \pythiaSF and
a C++ compiler for \pythiaE.
In addition, the database browser in \code{smodelsTools.py} requires IPython.

\smodels\ has been tested on Ubuntu, Scientific Linux (CERN) 5 and 6, as well
as on Mac OS X 10.11. The \smodelsnn\ package can be downloaded
from~\cite{smodels:wiki}. Unpacking the tarball with

\begin{Verbatim}[commandchars=\\\{\}]
tar -zxvf smodels-v1.1.x.tgz
\end{Verbatim}

\noindent
creates the directory \path{smodels-v1.1.x}, where the code
(subdirectory \allowbreak \path{smodels}) and the results database (subdirectory 
\path{smodels-database}) are located.

For installation, \smodelsnn\ makes use of Python's \code{setuptools}:

\begin{Verbatim}[commandchars=\\\{\}]
python setup.py install
\end{Verbatim}

\noindent
should install the entire project and compile the internal \pythia and
\nllfast versions.
It should also resolve the external dependencies, \ie\ install the Python
libraries listed above using e.g. \emph{pip}. If the Python libraries are
installed in a system folder (as is the default behavior), it will be necessary
to run the install command with superuser privilege. Alternatively, one can run
setup.py with the ``--user'' flag:

\begin{Verbatim}[commandchars=\\\{\}]
python setup.py install --user
\end{Verbatim}

In case the compilation of \smodelsnn\ fails, it is advised to try to compile the
tools manually, by issuing ``make'' in the \path{lib/} or in smodels-v1.1.x
directory. In case the installation of the external libraries fails, you can
also try to install them manually, then rerun \code{setup.py}.

To further help with installation and deployment, there is also a diagnostic
tool available:

\begin{Verbatim}[commandchars=\\\{\}]
python smodelsTools.py toolbox
\end{Verbatim}

\noindent
lists and checks all internal tools (\pythia and \nllfast) and external
(\code{numpy, scipy, unum, ...} ) dependencies.

More details on the installation procedure, in particular specific instructions
for installation on Ubuntu, Scientific Linux, other platforms or without
superuser privileges can be found in the \path{README.rst} file 
provided with the code.
In case everything fails, please contact the authors at 
\path{smodels-users@lists.oeaw.ac.at}.

The installation procedure also installs the \smodelsnn\ database of experimental results in the \code{smodels-database} subdirectory.
As mentioned in Section~\ref{Conclusions}, the v1.1.1 database release contains 186 results (125 upper limits and 61 efficiency maps)  
from 44 searches; 
24 of the EM-type results were `home-grown' using
\MA~\cite{Dumont:2014tja} and \CM~\cite{Drees:2013wra}
recasting, see Appendix~\ref{EMcreation}. 
The database also includes 35 preliminary results from 13 ATLAS and 4 CMS notes
which were superseded by a subsequent publication; they are kept in the
database for historical reasons but are not used in the default settings in \smodelsnn.
The complete list of analyses and results included in the database can be
consulted at~\cite{smodels:listofanalyses}.
We certify that all the results in the official database release have been
carefully validated and the validation material can be
found at~\cite{smodels:validation}.

\subsection{Adding Results to the Database}

The database can conveniently be updated independently from \smodelsnn\  code
updates. It suffices to unpack any new database tarball and replace the database
directory. In the same fashion, one can easily add additional results as
explained below.

\subsubsection{Adding FastLim data}
\label{Installation:adding-fastlim-data}

The official SModelS database can be augmented with data from the
FastLim~\cite{Papucci:2014rja} database. A tarball with the
\emph{properly converted fastlim-1.0} efficiency maps
can be downloaded from~
\cite{smodels:wiki}. 

Exploding the tarball in the top level directory of the database:
\begin{Verbatim}[commandchars=\\\{\}]
  mv smodels-v1.1-fastlim-1.0.tgz <smodels-database folder>
  cd <smodels-database folder>
  tar -xzvf smodels-v1.1-fastlim-1.0.tgz
  rm smodels-v1.1-fastlim-1.0.tgz
\end{Verbatim}
adds the FastLim-1.0 results in the standard directory structure (\ie\ in
\path{8TeV/ATLAS/}), see Section~\ref{DatabaseDefinitions:databasedefs}. 

\smodelsnn\ auto-detects FastLim results and issues an acknowledgement.
When using these results, please properly cite the FastLim paper~\cite{Papucci:2014rja} 
and the relevant experimental results; for convenience, a bibtex file is provided in the smodels-fastlim tarball.

For completeness we note that FastLim-1.0 contains 264 EM-type results 
based on 11 ATLAS conference notes, which were recast by the authors. 
Since in \smodels\ efficiencies with a relative statistical uncertainty greater than 25\% 
are set to zero and, moreover, zero-only EM's are discarded per default, effectively 
used in practice are 163 EM-type results from 9 conference notes.

\subsubsection{Adding one's own results}
\label{Installation:adding-one-s-own-results}

The database is organized as files in an ordinary directory hierarchy. Therefore,
adding additional experimental results is a matter of copying and editing text
files.
Once the new folders and files have been added following the 
format explained in Section~\ref{DatabaseDefinitions:databasedefs}, 
\smodelsnn\ automatically rebuilds the binary (pickle) database file.

\section{SModelS Tools}
\label{Tools::doc}\label{Tools:smodels-tools}

\smodelsnn\ comes with a number of tools that may be convenient for the user:

\begin{itemize}
    \item {}  a cross section calculator based on \pythia and \nllfast,

    \item {}  SLHA and LHE file checkers to check your input files for completeness 
		          and sanity,

    \item {}  a database browser to provide easy access to the database of experimental
    results,

\end{itemize}

\subsection{Cross-Section Calculator}
\label{Tools:xseccalc}\label{Tools:cross-section-calculator}

This tool computes LHC production cross sections for \emph{MSSM particles} and writes them out in an \emph{SLHA-type format} 
(see ``SLHA Format for Cross-Sections'' in Section~\ref{BasicInput:basicinput})
This can be particularly convenient for adding cross sections to SLHA input files.
The calculation is done at LO with \pythia (optionally \pythiaSF
or \pythiaET); K-factors for colored particles are computed with \nllfast. \\

\noindent 
\textbf{The usage of the cross section calculator is:}

\begin{Verbatim}[frame=lines]
smodelsTools.py xseccomputer [-h] -f FILENAME 
                [-s SQRTS [SQRTS ...]] [-e NEVENTS] [-v VERBOSITY] 
                [-c NCPUS] [-p] [-q] [-k] [-n] [-N] [-O] [-6] [-8]
\end{Verbatim}

\begin{description}
    \item[{\emph{arguments}:}] \leavevmode\begin{optionlist}{2cm}
    \item [-h, -{-}help]  show help message and exit.
    \item [-s SQRTS, -{-}sqrts SQRTS] (int): LHC center-of-mass energy in TeV for computing the cross sections. Can be more than one value; default is both 8 and 13.
    \item [-e NEVENTS, -{-}nevents NEVENTS] (int): number of Monte Carlo
events to be simulated when running \pythia.
    \item [-c NCPUS, -{-}ncpus NCPUS] (int):   number of CPU cores to be used simultaneously. $-1$ means `all'. This is only used
    when cross sections are computed for multiple SLHA files.
    \item [-p, -{-}tofile] if set, the cross sections will be written back to the
    file. If the input file already contains cross sections, only new records
    will be written. If not set, the cross sections will be
    written to the screen, only.
    \item [-q, -{-}query]  only query whether the input file already contains cross sections.
    \item [-k, -{-}keep]   keep the temporary directory containing the
    \pythia run output. This option is only relevant when checking for errors
    when running \pythia.
    \item [-n, -{-}NLO]  use \pythia and \nllfast to compute NLO
    cross sections (default is LO). Note that since \nllfast only contains results for production
    of gluinos and squarks (incl.\ 3rd generation), only these cross sections will be generated.
    \item [-N, -{-}NLL]  use \pythia and \nllfast to compute NLO+NLL
    cross sections (takes precedence over NLO, default is LO). Note that since NLLfast only contains results for production
    of gluinos and squarks (incl.\ 3rd generation), only these cross sections will be generated.
    \item [-O, -{-}LOfromSLHA]  if set, SModelS will read the LO cross sections from
    the input file and use NLLfast to compute the NLO or NLO+NLL cross sections
    for squarks and gluinos. 
    \item [-f FILENAME, -{-}filename FILENAME]  SLHA file to compute cross sections for. If a directory is given, compute cross sections for all files in the directory.
    \item [-v VERBOSITY, -{-}verbosity VERBOSITY] Verbosity (`debug', `info',
`warning', `error'. Default is `info').
    \item [-6, -{-}pythia6] use \pythiaS for LO cross sections.
    \item [-8, -{-}pythia8] use \pythiaE for LO cross sections (default).

\end{optionlist}

\end{description}

\noindent
Further \pythia parameters are defined in \code{smodels/etc/pythia8.cfg} \\
(\pythiaE) or \code{smodels/etc/pythia.card} (\pythiaS).\\

\noindent
A typical usage example is:

\begin{Verbatim}[commandchars=\\\{\},frame=none]
smodelsTools.py xseccomputer \PYGZhy{}s 8 13 \PYGZhy{}e 50000 \PYGZhy{}p 
                \PYGZhy{}f inputFiles/slha/compressedSpec.slha -6 
\end{Verbatim}

\noindent
which will compute 8~TeV and 13~TeV LO cross sections (at the LHC) with 
\pythiaS for all MSSM
processes using 50k MC events. If, \emph{after} the LO cross sections have been
computed, one wants to add the NLO+NLL cross sections for gluinos and squarks (incl.\ 3rd generation):

\begin{Verbatim}[commandchars=\\\{\},frame=none]
smodelsTools.py xseccomputer -s 8 13 -p -N -O 
                -f inputFiles/slha/compressedSpec.slha
\end{Verbatim}

The resulting file will then contain LO cross sections for all MSSM processes
and NLO+NLL cross sections for the available processes in
NLLfast (gluino and squark production).\footnote{If a higher precision is needed, for instance for electroweak production or 
for gluino/squark masses outside the \nllfast\ grids, 
this has to be taken care of by the user.}
When reading the input file, SModelS will then
use only the highest-order cross section available for each process.

\subsection{Input File Checks}
\label{Tools:input-file-checks}\label{Tools:filechecks}

As discussed in Section~\ref{BasicInput:basicinput}, SModelS accepts both SLHA and LHE input files. It can be convenient to perform certain
sanity checks on these files as described below.

\paragraph{\bf LHE File Checker}
\label{Tools:lhe-file-checker}\label{Tools:lhechecks}

For an LHE input file only very basic checks are performed, namely that

\begin{itemize}
    \item {}  the file exists,

    \item {}  it contains at least one event,

    \item {}  the information on the total cross section and the center of mass energy can be found.
\end{itemize}

\noindent
\textbf{The usage of the LHE checker is simply:}

\begin{Verbatim}[commandchars=\\\{\},frame=lines]
smodelsTools.py lhechecker [-h] -f FILENAME
\end{Verbatim}

\begin{description}
    \item[{\emph{arguments}:}] \leavevmode\begin{optionlist}{3cm}
\item [-h, -{-}help]  
show this help message and exit
\item [-f FILENAME, -{-}filename FILENAME]  
name of input LHE file
\end{optionlist}
\end{description}

\noindent
A typical usage example is:

\begin{Verbatim}[commandchars=\\\{\},frame=none]
smodelsTools.py lhechecker -f 
                inputFiles/slha/gluino_squarks.lhe
\end{Verbatim}

\paragraph{\bf SLHA File Checker}
\label{Tools:slha-file-checker}\label{Tools:slhachecks}
The SLHA file checker allows to perform quite rigorous checks of SLHA input
files. Concretely, it verifies that

\begin{itemize}
    \item {}  the file exists and is given in SLHA format,

    \item {}  the file contains masses and decay branching ratios in standard
    SLHA format,

    \item {}  the file contains cross sections according to the
    SLHA format for cross sections,

    \item {}  the lightest $\mathbb{Z}_2$\emph{-odd state} (the
    LSP in supersymmetric models) is neutral,

    \item {} there are no stable charged particles nor displaced vertices (no
    non-prompt visible decays), as currently all the analyses considered by
    SModelS require a prompt MET signature.
\end{itemize}

\noindent
In addition, one can ask that
\begin{itemize}
\item {} 
all decays listed in the DECAY block are kinematically allowed, \emph{i.e.} the
sum of masses of the decay products may not exceed the mother mass. \emph{This
check for ``illegal decays'' is turned off by default.}
\end{itemize}

\noindent
If any of the above tests fails (returns a negative result), an error message is shown.

Some more comments are in order.
In order to check that the lightest $\mathbb{Z}_2$-odd state has zero electric and color charges, the quantum numbers of the BSM particles must be given in the
\code{qNumbers} dictionary in \code{particles.py}. The format is\\

\noindent
{\tt [2*spin, 3*electric charge, dimension of SU(3) representation]}\\

\noindent
The list of quantum numbers is also required to check for displaced vertices or
heavy charged particles.
The check for long-lived (or stable) particles first verifies that these appear in
one of the cross section blocks and their cross section exceeds the minimum
cross section value defined by \code{sigmacut} in the parameters file, see Section~\ref{RunningSModelS:parameterfile}.
If the cross section is larger than \code{sigmacut} and the particle is stable, the
checker checks if it is neutral (both electric and color charges are zero). On
the other hand, if the particle is unstable, but its lifetime (times \emph{c})
is larger than a minimum value (\emph{default = 10 mm}), the particle is
considered a non-prompt decay. For these cases all channels are then
checked for visible decay products. If the branching ratio to visible decays
times the maximum production cross section for the particle exceeds
\code{sigmacut}, the particle's decay is considered a displaced vertex. \\

\noindent
\textbf{The usage of the SLHA checker is:}

\begin{Verbatim}[commandchars=\\\{\},frame=lines]
smodelsTools.py slhachecker [-h] [-xS] [-lsp] [-longlived] 
                [-m DISPLACEMENT] [-s SIGMACUT] [-illegal] 
                [-dB] -f FILENAME
\end{Verbatim}

\begin{description}
    \item[{\emph{arguments}:}] \leavevmode\begin{optionlist}{3cm}
    \item [-h, -{-}help]  show this help message and exit
    \item [-xS, -{-}xsec]  turn off the check for xsection blocks
    \item [-lsp, -{-}lsp]  turn off the check for charged lsp
    \item [-longlived, -{-}longlived]  turn off the check for stable charged particles and visible displaced vertices
    \item [-m DISPLACEMENT, -{-}displacement DISPLACEMENT]  give maximum displacement of secondary vertex in m
    \item [-s SIGMACUT, -{-}sigmacut SIGMACUT] give sigmacut in fb
    \item [-illegal, -{-}illegal]  turn on check for kinematically forbidden decays
    \item [-dB, -{-}decayBlocks]  turn off the check for missing decay blocks
    \item [-f FILENAME, -{-}filename FILENAME]  name of input SLHA file
\end{optionlist}
\end{description}

\noindent
A typical usage example is:

\begin{Verbatim}[commandchars=\\\{\},frame=none]
smodelsTools.py slhachecker -m 0.001 -s 0.01 
                -f inputFiles/slha/lightSquarks.slha
\end{Verbatim}

\noindent
Running this will print the status flag and a message with potential warnings
and error messages.

\subsection{Database Browser}
\label{Tools:database-browser}\label{Tools:databasebrowser}

The database browser is a tool based on IPython which provides an easy way to directly access
the SModelS database. It owns several methods to select  experimental results or data sets
satisfying some user-defined conditions as well as to access the meta data and
data inside each experimental result. \\

\noindent
\textbf{The usage of the browser interface is:}

\begin{Verbatim}[commandchars=\\\{\},frame=lines]
smodelsTools.py database-browser [-h] -p PATH_TO_DATABASE [-t]
\end{Verbatim}

\begin{description}
    \item[{\emph{arguments}:}] \leavevmode\begin{optionlist}{3cm}
    \item [-h, -{-}help]   show this help message and exit
    \item [-p PATH\_TO\_DATABASE,]
    \item [-{-}path\_to\_database PATH\_TO\_DATABASE]   path to SModelS database
    \item [-t, -{-}text]   load text database, don't even search for binary database file
\end{optionlist}

\end{description}

\noindent
A typical usage example is:

\begin{Verbatim}[commandchars=\\\{\},frame=none]
smodelsTools.py database\PYGZhy{}browser \PYGZhy{}p ./smodels\PYGZhy{}database
\end{Verbatim}

Loading the database may take a few seconds if the binary database file exists. Otherwise the {pickle file} will be created. Starting the browser opens an IPython session, which can be used to select specific experimental results (or groups of experimental results),
check upper limits and/or efficiencies for specific masses/topologies and access all the available information in the database. A simple example is given below:

\begin{Verbatim}[commandchars=\\\{\},frame=none,fontsize=\small]
In [1]: print browser  \PYGZsh{}Print all experimental results in the browser
[\PYGZsq{}ATLAS\PYGZhy{}SUSY\PYGZhy{}2015\PYGZhy{}09\PYGZsq{}, \PYGZsq{}CMS\PYGZhy{}SUS\PYGZhy{}PAS\PYGZhy{}15\PYGZhy{}002\PYGZsq{}, \PYGZsq{}ATLAS\PYGZhy{}CONF\PYGZhy{}2012\PYGZhy{}105\PYGZsq{},  
...]

In [2]: browser.selectExpResultsWith(txName = \PYGZsq{}T1tttt\PYGZsq{}, dataType = 
\PYGZsq{}upperLimit\PYGZsq{}) \PYGZsh{}Select only the UL results with the topology T1tttt

In [3]: print browser \PYGZsh{}Print all experimental results in the browser 
(after selection)
[\PYGZsq{}ATLAS\PYGZhy{}SUSY\PYGZhy{}2015\PYGZhy{}09\PYGZsq{}, \PYGZsq{}CMS\PYGZhy{}SUS\PYGZhy{}PAS\PYGZhy{}15\PYGZhy{}002\PYGZsq{}, \PYGZsq{}ATLAS\PYGZhy{}CONF\PYGZhy{}2012\PYGZhy{}105\PYGZsq{}, 
\PYGZsq{}ATLAS\PYGZhy{}CONF\PYGZhy{}2013\PYGZhy{}007\PYGZsq{}, \PYGZsq{}ATLAS\PYGZhy{}CONF\PYGZhy{}2013\PYGZhy{}061\PYGZsq{}, \PYGZsq{}ATLAS\PYGZhy{}SUSY\PYGZhy{}2013\PYGZhy{}04\PYGZsq{}, 
\PYGZsq{}ATLAS\PYGZhy{}SUSY\PYGZhy{}2013\PYGZhy{}09\PYGZsq{}, \PYGZsq{}ATLAS\PYGZhy{}SUSY\PYGZhy{}2013\PYGZhy{}18\PYGZsq{}, \PYGZsq{}CMS\PYGZhy{}PAS\PYGZhy{}SUS\PYGZhy{}12\PYGZhy{}026\PYGZsq{}, 
...]

\PYGZsh{}Define masses for the T1tttt topology:
In [4]: gluinoMass, LSPmass = 800.*GeV, 100.*GeV 

\PYGZsh{}Get UL for a specific experimental result
In [5]: browser.getULFor(\PYGZsq{}CMS\PYGZhy{}SUS\PYGZhy{}PAS\PYGZhy{}15\PYGZhy{}002\PYGZsq{},\PYGZsq{}T1tttt\PYGZsq{},
[[gluinoMass,LSPmass],[gluinoMass,LSPmass]]) 
Out[5]: 5.03E\PYGZhy{}02 [pb]

\PYGZsh{}Get the upper limits for all the selected results for the given
\PYGZsh{}topology and mass
In [6]: for expResult in browser:  
   ...:     print expResult.getValuesFor(\PYGZsq{}id\PYGZsq{}),\PYGZsq{}UL = \PYGZsq{},
   expResult.getUpperLimitFor(txname=\PYGZsq{}T1tttt\PYGZsq{},mass=
   [[gluinoMass,LSPmass],[gluinoMass,LSPmass]])
   ...:
[\PYGZsq{}ATLAS\PYGZhy{}SUSY\PYGZhy{}2015\PYGZhy{}09\PYGZsq{}] UL =  None
[\PYGZsq{}CMS\PYGZhy{}SUS\PYGZhy{}PAS\PYGZhy{}15\PYGZhy{}002\PYGZsq{}] UL =  5.03E\PYGZhy{}02 [pb]
[\PYGZsq{}ATLAS\PYGZhy{}CONF\PYGZhy{}2012\PYGZhy{}105\PYGZsq{}] UL =  6.70E\PYGZhy{}02 [pb]
[\PYGZsq{}ATLAS\PYGZhy{}CONF\PYGZhy{}2013\PYGZhy{}007\PYGZsq{}] UL =  2.40E\PYGZhy{}02 [pb]
 ...
 
\PYGZsh{}Print the luminosities for the selected experimental results 
In [7]: for exp in browser:  
   ...:     print exp.getValuesFor(\PYGZsq{}id\PYGZsq{}), exp.getValuesFor(\PYGZsq{}lumi\PYGZsq{})
   ...:
[\PYGZsq{}ATLAS\PYGZhy{}SUSY\PYGZhy{}2015\PYGZhy{}09\PYGZsq{}] [3.20E+00 [1/fb]]
[\PYGZsq{}CMS\PYGZhy{}SUS\PYGZhy{}PAS\PYGZhy{}15\PYGZhy{}002\PYGZsq{}] [2.20E+00 [1/fb]]
[\PYGZsq{}ATLAS\PYGZhy{}CONF\PYGZhy{}2012\PYGZhy{}105\PYGZsq{}] [5.80E+00 [1/fb]]
[\PYGZsq{}ATLAS\PYGZhy{}CONF\PYGZhy{}2013\PYGZhy{}007\PYGZsq{}] [2.07E+01 [1/fb]]
...
\end{Verbatim}

%

\clearpage
\section{Home-grown Efficiency Maps  \label{EMcreation}}

While most efficiency maps included in the \smodels\ database were directly provided by the experimental collaborations as auxiliary material  for their publications, we also produced a number of them ourselves in order to improve the coverage of simplified models. Such `home-grown' efficiency maps are particularly relevant for topologies with one or more intermediate particles: here we need several mass planes for the interpolation, but often only one is provided in the official results. The home-grown efficiency maps included in the v1.1.1 database are summarised in Table~\ref{tab:processdetails}. The TxNames denote  the following simplified models:\footnote{For gluinos decays, charge-conjugated final states are assumed implicitly.} 

\begin{table}[t!]\centering
\footnotesize
 \begin{tabular}{|l|l|l|} \hline
Analysis & Process & SMS topologies  \\ \hline\hline
\multirow{2}{*}{ATLAS-SUSY-2013-04~\cite{Aad:2013wta}}
        &  $pp \to \tilde g \tilde g \,(j)$ & T1bbbb, T1btbt, T5,
T5WW, T5ZZ   \\ 
        &  $pp \to \tilde t_1 \tilde t_1^* \,(j)$  & T2tt,T6bbWW  \\ \hline
\multirow{2}{*}{CMS-SUS-13-012~\cite{Chatrchyan:2014lfa}}
        & $pp \to \tilde g\tilde g \,(j)$ & T1bbbb,  T1btbt, T5,
T5bbbb, T5tttt, T5WW, T5ZZ \\
        &  $pp \to \tilde t_1 \tilde t_1^* \, (j)$  & T2tt, T6bbWW  \\
        & $pp \to \tilde b_1 \tilde b_1^* \, (j)$  & T2bb  \\
        & $ pp \to \tilde\chi^\pm_1 \tilde\chi^0_2 \,(j)$ & TChiWZ   \\
        & $ pp \to \tilde\chi^+_1 \tilde\chi^-_1\,(j)$ & TChiWW   \\
        & $ pp \to \tilde\chi^0_2 \tilde\chi^0_2\,(j)$ & TChiZZ   \\ \hline
\multirow{2}{*}{ATLAS-SUSY-2013-11~\cite{Aad:2014vma}}
        & $ pp \rightarrow  \tilde{l} \tilde{\bar l}$ &  TSlepSlep \\
        & $ pp \to \tilde\chi^+_1 \tilde\chi^-_1$ & TChiWW,
TChipChimSlepSnu   \\ \hline
ATLAS-SUSY-2013-05~\cite{Aad:2013ija}  & $pp \to \tilde b_1 \tilde
b_1^* (j)$ & T2bb   \\[1mm] \hline
 \end{tabular}
  \caption{Summary of the `home-grown' efficiency maps included in the
v1.1.1 database.}
  \label{tab:processdetails}
\end{table}

\noindent 
\begin{itemize}
\item T1bbbb:  $p p \rightarrow \tilde g \tilde g$, $ \tilde g \rightarrow b \bar b \tilde\chi^0_1$\,;
\item T1btbt:  $p p \rightarrow \tilde g \tilde g$, $ \tilde g \rightarrow b t \tilde\chi^0_1$\,;
\item T2bb:  $p p \rightarrow \tilde b_1 \tilde b_1^*$, $\tilde b_1 \rightarrow b \tilde \chi^0_1$\,;
\item T2tt: $p p \rightarrow \tilde t_1 \tilde t_1^*$, $\tilde t_1 \rightarrow t \tilde \chi^0_1$\,;
\item T5:  $p p \rightarrow \tilde g \tilde g$, $ \tilde g \rightarrow q \tilde q, \tilde q \rightarrow q \tilde\chi^0_1$\,; 
\item T5bbbb:  $p p \rightarrow \tilde g \tilde g$, $\tilde g \rightarrow b\tilde{b}_1 , \tilde b_1 \rightarrow b \tilde\chi^0_1$\,;
\item T5tttt:  $p p \rightarrow \tilde g \tilde g$, $\tilde g \rightarrow t\tilde{t}_1 , \tilde t_1 \rightarrow t \tilde\chi^0_1$\,;
\item T5WW:  $p p \rightarrow \tilde g \tilde g$, $\tilde g \rightarrow q \bar{q}' \tilde \chi^{\pm}_1$, $\tilde \chi^{\pm}_1 \rightarrow W^\pm \tilde \chi^0_1$\,;
\item T5ZZ:  $p p \rightarrow \tilde g \tilde g$, $\tilde g \rightarrow q \bar{q} \tilde \chi^0_2$, $\tilde \chi^0_2 \rightarrow Z \tilde \chi^0_1$\,;
\item T6bbWW:  $p p \rightarrow \tilde t_1 \tilde t_1^*$, $\tilde t_1 \rightarrow b \tilde\chi^+_1$, $\tilde \chi^+_1 \rightarrow W^+ \tilde \chi^0_1$\,;
\item TChiWW: $p p \rightarrow \tilde \chi^+_1 \tilde \chi^-_1$, $\tilde \chi^\pm_1 \rightarrow W^{\pm} \tilde \chi^0_1$\,;
\item TChiWZ: $p p \rightarrow  \tilde \chi^\pm_1 \tilde \chi^0_2$, $\tilde\chi^\pm_1 \rightarrow W^{\pm} \tilde \chi^0_1$, $\tilde \chi^0_2 \rightarrow Z \tilde \chi^0_1$  \,;
\item TChiZZ: $p p \rightarrow \tilde \chi^0_2 \tilde \chi^0_2 $, $\tilde \chi^0_2 \rightarrow Z \tilde \chi^0_1$\,; 
\item TSlepSlep:  $p p \rightarrow \tilde e^+ \tilde e^-$ or $\tilde \mu ^+ \tilde \mu^-$, $ \tilde e^{\pm} \rightarrow e^{\pm} \tilde\chi^0_1$, $ \tilde \mu^{\pm} \rightarrow \mu^{\pm} \tilde\chi^0_1$\,;
\item TChipChimSlepSnu: $p p \rightarrow \tilde \chi^+_1 \tilde \chi^-_1$,  
    $\tilde{\chi}_1 ^{\pm}  \rightarrow l^{\pm}\tilde{\nu}_{l} $ or $\nu_{l}\tilde{l}^{\pm}$,   
    $\tilde{\nu}_l \rightarrow \nu_l \tilde{\chi}_1^0$, $\tilde{l}^{\pm} \rightarrow  l^{\pm}\tilde{\chi}_1^0$\,.
\end{itemize}

Monte Carlo events for the above processes were generated using \amc~\cite{Alwall:2014hca} (v.2.3.3) with decay and
hadronisation performed via the integrated installation of {\sc Pythia\,6.4}~\cite{Sjostrand:2006za}. 
For recasting analyses sensitive to hadronic final states, events were generated including up to one extra jet, 
see Table \ref{tab:processdetails}.  The merging was performed
using Matrix Element--Parton Shower (ME-PS) matching 
according to the $k_T$-jet MLM scheme~\cite{MLM:scheme,Hoche:2006ph,Alwall:2007fs} with merging and matching parameters set as $(Qcut,XQcut)=(90,50)$~GeV, following CMS practice.
The hadronised samples were then passed through \MA~\cite{Dumont:2014tja,Conte:2014zja} (v1.4.4) or \CM~\cite{Drees:2013wra} (v1.2.2) 
for further processing. Both these codes interface to the \delphes~\cite{deFavereau:2013fsa} detector simulation, 
using {\sc FastJet}~\cite{Cacciari:2011ma} for jet clustering. Concretely, we used the following recasting codes:

\noindent 
\begin{itemize}
\item \MA\ recast code \cite{ATLAS-SUSY-2013-04MA5} for the ATLAS multi-jet + MET analysis~\cite{Aad:2013wta}; \
\item \MA\ recast code \cite{CMS-SUS-13-012MA5} for the CMS multi-jet + MET analysis~\cite{Chatrchyan:2014lfa}; \
\item \MA\ recast code \cite{ATLAS-SUSY-2013-11MA5} for the ATLAS dilepton plus MET analysis~\cite{Aad:2014vma}; 
\item \CM~\cite{Drees:2013wra} for the ATLAS third generation squark search~\cite{Aad:2013ija}. \\
\end{itemize}

In the case of topologies with one-step cascade decays (T5, T5bbbb, T5tttt, T5WW, T5ZZ, T6bbWW, TChipChimSlepSnu) we created at least three different mass planes, so that the mass of the intermediate sparticle assumes a value close to the mother mass, close to the neutralino mass and equally spaced between the two, in order to obtain a good coverage of the parameter space. To this end, a parameterization in terms of relative mass differences, 
$m_{Interm.} = x\cdot m_{Mother} + (1 - x) \cdot m_{\tilde \chi^0_1}$ with $x={\rm const.}$, was used.  
Moreover, 
where relevant, we produced planes with fixed mass gaps of $\Delta M(A,B)=m_A-m_B$ 
to improve the accuracy of the interpolation, \eg\
for T5tttt near the kinematic boundary for on-shell top quarks, and for the T5WW and T6bbWW models (only for CMS-SUS-13-012)
in the region where the chargino decays via an off-shell $W$ boson.
For the T1btbt model, the chargino and the LSP were considered mass-degenerate.
An overview of the various mass planes is given in Table~\ref{tab:massplanes}.


\begin{table}\centering\footnotesize
\begin{tabular}{| l | l | l |}
\hline
TxName & mass planes, $x$ parametrization & mass planes, fixed $\Delta M(A,B)$ [GeV] \\ \hline \hline
T5		&  $x= 0.05,\, 0.5,\, 0.95$ &  \\
T5tttt     	& $x=0.5$ & $\Delta M(\tilde g,\tilde t_1) = 177$, $\Delta M(\tilde t_1,\tilde\chi^0_1) = 177$ \\
T5bbbb 	& $x = 0.05,\, 0.5,\, 0.95$ & \\
T5WW     	& $x = 0.05,\, 0.5,\, 0.95$ & $\Delta M(\tilde\chi^\pm_1,\tilde\chi^0_1) = 10, 75$ \\
T5ZZ	& $x = 0.05,\, 0.5,\, 0.95$ & \\
T6bbWW 	& $x = 0.1,\, 0.5,\, 0.9$ & $\Delta M(\tilde\chi^\pm_1,\tilde\chi^0_1) = 10, 75$ \\
TChipChimSlepSnu & $x=0.05,\, 0.25,\, 0.50,\, 0.75,\, 0.95$ & $\Delta M(\tilde\chi^\pm_1,\tilde l)=5,10,15$, $\Delta M(\tilde l,\tilde\chi^0_1)=5,10,15$ \\
\hline
\end{tabular}
\caption{Mass planes produced for topologies with one-step cascade decays. \label{tab:massplanes}}
\end{table}

In the input SLHA files, we assumed diagonal mixing matrices for both charginos and neutralinos, giving a pure bino $\tilde \chi^0_1$ and pure winos $\tilde \chi^{\pm}_1$ and $\tilde \chi^0_2$. However, since  {\sc Pythia\,6.4} was used for decaying the intermediate particles, any effects in kinematic distributions that might arise from sparticle mixing (due to polarization of the decay products) as well as spin correlations are disregarded. 
No additional radiation was produced for ATLAS-SUSY-2013-11 since the signal regions targeting these specific SMS veto the presence of jets.
For the TSlepSlep simplified model, we took $\tilde l \equiv \tilde e_L, \tilde e_R, \tilde\mu_L, \tilde\mu_R$, with all of them being mass-degenerate. For the TChipChimSlepSnu model, the intermediate left-handed sleptons and sneutrinos are taken mass degenerate for all the three slepton flavours. 

We have verified that our procedure reproduces well the official 95\%~CL exclusion curves whenever provided by the experimental collaborations; the validation plots are available at~\cite{smodels:validation}. For completeness, we also note that detailed validation notes for the \MA\ recast codes are available at~\cite{ATLASSUSY201304MA5,CMSSUS13012validation,ATLASSUSY201311validation} and for  \CM\ at~\cite{CMvalidation}.

\clearpage
\bibliography{references}

\end{document}